%% file: paper.tex
\documentclass[journal]{vgtc}                % final (journal style)
\ifpdf%                                % if we use pdflatex
  \pdfoutput=1\relax                   % create PDFs from pdfLaTeX
  \usepackage{graphicx}                % allow us to embed graphics files
  \DeclareGraphicsExtensions{.pdf,.png,.jpg,.jpeg} % for pdflatex we expect .pdf, .png, or .jpg files
\else%                                 % else we use pure latex
  \usepackage{graphicx}                % allow us to embed graphics files
  \DeclareGraphicsExtensions{.eps}     % for pure latex we expect eps files
\fi%

\graphicspath{{figures/}{pictures/}{images/}{./}} % where to search for the images

\usepackage{microtype}                 % use micro-typography (slightly more compact, better to read)
\PassOptionsToPackage{warn}{textcomp}  % to address font issues with \textrightarrow
\usepackage{textcomp}                  % use better special symbols
\usepackage{mathptmx}                  % use matching math font
\usepackage{times}                     % we use Times as the main font
\usepackage{cite}                      % needed to automatically sort the references
\usepackage{tabu}                      % only used for the table example
\usepackage{booktabs}                  % only used for the table example
\newcommand{\toolname}{Kori}

\newcolumntype{L}{>{\raggedright\arraybackslash}X}
\newcolumntype{C}{>{\centering\arraybackslash}X}
\newcolumntype{R}{>{\raggedleft\arraybackslash}X} 
\newcommand{\bpstart}[1]{\vspace{1mm} \noindent{\textbf{#1.}}}

\usepackage{amssymb}
\usepackage{tikz}
\usepackage{xcolor}

\definecolor{myblue}{RGB}{0,162,255}

\newcommand*\circleBlue[1]{\tikz[baseline=(char.base)]{
            \node[shape=circle,fill=myblue, inner sep=1pt] (char) {\textcolor{white}{\small{#1}}};}}
            
\newcommand*\rectBlue[1]{\tikz[baseline=(char.base)]{
            \node[shape=rectangle,fill=myblue, inner sep=2pt] (char) {\textcolor{white}{\small{#1}}};}}
            
\usepackage[T1]{fontenc}
\usepackage[scaled=.85]{inconsolata}

\newcommand{\inlinegraphic}[1]{%
	\protect\raisebox{-2.6pt}{%
		\protect\includegraphics[height=0.37cm]{#1}%
	}\,%
}

\newcommand{\inlineboxplot}[1]{%
	\protect\raisebox{-2pt}{%
		\protect\includegraphics[height=0.28cm, trim=0.6in 0.2in 0.5in 0in, clip=true]{#1}%
	}\,%
}

\onlineid{0}

\vgtccategory{Research}
\vgtcpapertype{please specify}

\title{Kori: Interactive Synthesis of Text and Charts in Data Documents}

\author{Submission ID 1008}
 \author{Shahid Latif, Zheng Zhou, Yoon Kim, Fabian Beck, and Nam Wook Kim}
  \authorfooter{
  \item
   Shahid Latif and Fabian Beck are with University of Duisburg-Essen. E-mail: \{shahid.latif, fabian.beck\}@uni-due.de.
  \item
   Zheng Zhou and Nam Wook Kim are with Boston College. E-mail: \{zheng.zhou, nam.wook.kim\}@bc.edu.
  \item
   Yoon Kim is with Harvard University. E-mail: yoonkim@g.harvard.edu.
  }

\shortauthortitle{Latif \MakeLowercase{\textit{et al.}}: Kori: Interactive Synthesis of Text and Charts in Data Documents}
\abstract{Charts go hand in hand with text to communicate complex data and are widely adopted in news articles, online blogs, and academic papers. They provide graphical summaries of the data, while text explains the message and context. However, synthesizing information across text and charts is difficult; it requires readers to frequently shift their attention. We investigated ways to support the tight coupling of text and charts in data documents. To understand their interplay, we analyzed the design space of chart--text references through news articles and scientific papers. Informed by the analysis, we developed a mixed-initiative interface enabling users to construct interactive references between text and charts. It leverages natural language processing to automatically suggest references as well as allows users to manually construct other references effortlessly. A user study complemented with algorithmic evaluation of the system suggests that the interface provides an effective way to compose interactive data documents.%
} % end of abstract

\teaser{
  \centering
  \includegraphics[width=\linewidth, trim=0in 3.1in 2.0in 0.2in, clip=true]{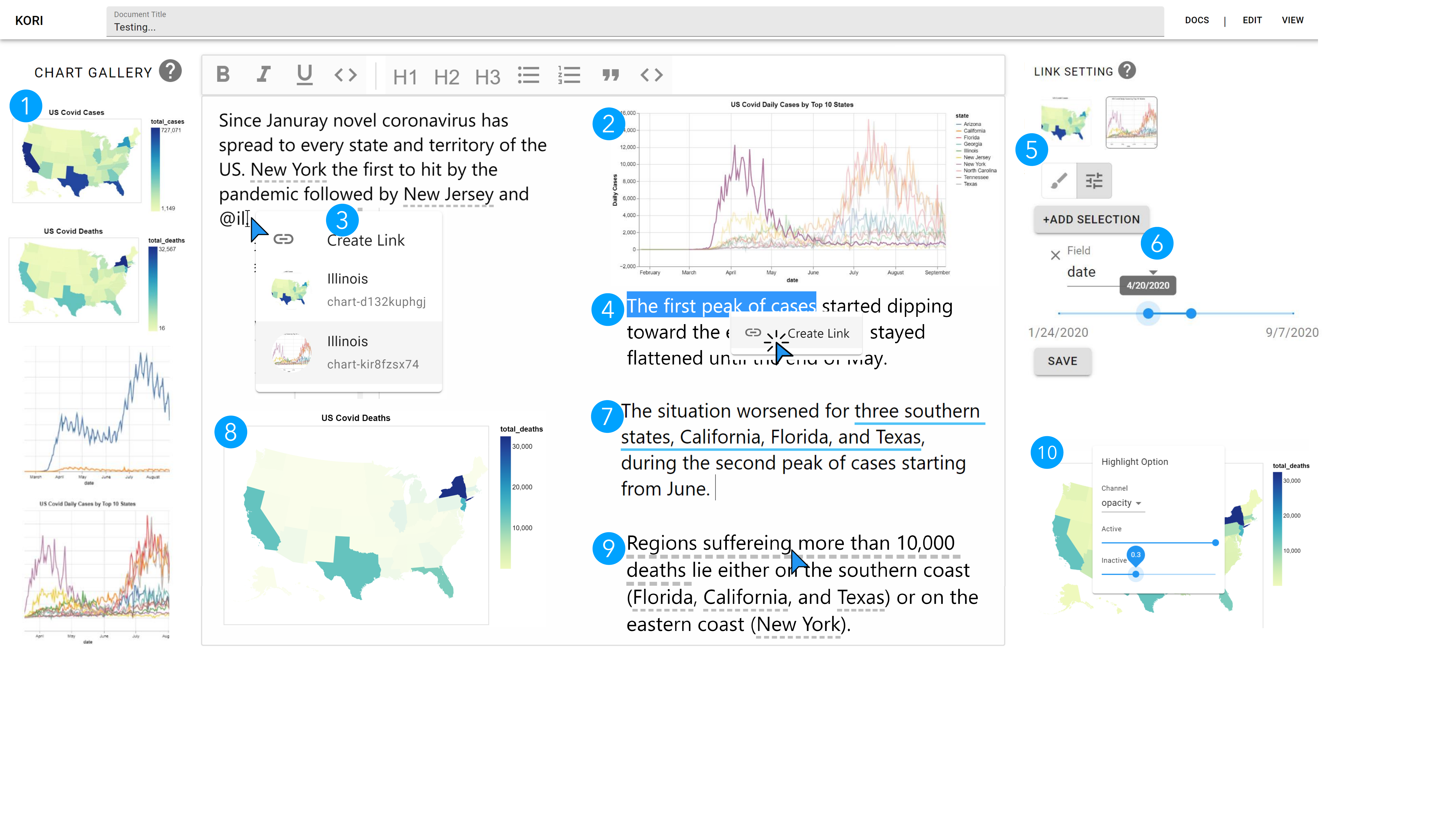}
  \caption{The \toolname{} system consists of a chart gallery (left), edit area (middle), and a link setting panel (right). Kori automatically suggests potential references (dotted gray underline) as a user types. Besides, it supports manual creation of links through simple interactions (\textbf{4}--\textbf{6}). The steps \textbf{1}--\textbf{10} describe a usage scenario to create an interactive story (see \autoref{sec:usage}).}
  \label{fig:teaser}
}

\vgtcinsertpkg

\begin{document}


\firstsection{Introduction}
\maketitle

%% \section{Introduction} %for journal use above \firstsection{..} instead
As our society becomes more data-driven, the need for communicating data to a broader audience is inevitable. This has led to the emergence of narrative visualization and data-driven storytelling~\cite{riche2018data}, going beyond traditional analysis-focused visualization systems. Charts are now frequently adopted in various forms of data stories such as news magazines, slide shows, and videos~\cite{segel2010narrative}. They are integrated with explanatory text to produce a synergetic effect. Charts can engage an audience and present a perceptually effective representation of data, while the text provides guidance to readers and explains the additional context. Together, charts and text are essential for telling compelling stories and effectively conveying messages with data.

However, it is often challenging to synthesize information across two distinct media---the text and charts---as they are spatially separated apart~\cite{kong2014extracting}. The readers have to switch their attention back and forth to find references between text and visual marks on the charts that encode data values (e.g., bars, lines, points) and vice versa. This phenomenon is known as split-attention effect~\cite{ayres2005split} in cognitive load theory~\cite{sweller2005implications}---or the contiguity principle in cognitive theory of multimedia learning~\cite{mayer201412}---that incurs a significant cognitive burden on readers' working memory and can have a negative impact on learning~\cite{ayres2012split,castro2019instructional}. Several recent works have begun to explore ways of resolving this issue by demonstrating an explicit interactive linking between text and associated parts of charts~\cite{kong2014extracting,zhi2019linking,lalle2019gaze} or through the use of signaling principle to incorporate visual cues to guide attention~\cite{mayer201412}. However, establishing and supporting such interactive references requires advanced programming knowledge that is mostly absent from authors of data-driven articles. Authoring tools for visualization and data storytelling~\cite{kim2019datatoon,satyanarayan2014authoring,gratzl2016visual} do not currently support the construction of such references.

In this work, we investigate ways to support the production of tight coupling between text and charts. To better understand their interplay, we first analyzed text--chart pairs from existing sources. We expanded a collection from news media outlets containing bar charts~\cite{kong2014extracting} with additional chart types, including line charts, scatter plots, pie charts, and maps, and widened the scope to also include examples from scientific journals. We found that text--chart references have similarities to selection operations in a chart (e.g., point and interval selections). The textual phrases that refer to such selections can be aggregated in a hierarchical manner, similar to a semantic parse tree, and result in a \textit{union} or \textit{intersection} of the visual marks in the chart.

Based on the design space analysis of text--chart references, we developed \toolname{}, a mixed-initiative interface that enables the synthesis of text and charts through interactive references; \autoref{fig:teaser} shows the interface. It supports three main interactions: (i) accepting or rejecting suggested references, (ii) explicitly triggering suggestions \circleBlue{3}, and (iii) manually constructing references using simple interactions \circleBlue{4} -- \circleBlue{6}. While authoring, \toolname{} automatically identifies and suggests references \circleBlue{9}. Our automatic suggestion approach leverages natural language processing to derive point- and interval-level references. Parsed dependencies group these references to generate higher-level references if two or more of them are semantically related.

Our evaluation of \toolname{} is two-fold. First, we quantitatively evaluated the automatic reference suggestion approach. %The results highlight several limitations such as , suggesting an exciting avenue for future research. 
Second, we qualitatively evaluated the interface through a user study. We recruited eleven participants, including visualization experts, novices, and interface designers, and asked them to perform three tasks of reproducing and creating text--chart references, followed by a usability survey and a reflection interview. %Third, we evaluated the system along cognitive dimensions of notations. 
The results suggest that \toolname{} provides a novel, intuitive, and easy-to-use interface to write an interactive data-driven document, that has the potential to advance the current practice of existing authoring tools. %We also found that automatic reference suggestions works well and complements the manual construction of references.

\section{Related Work}

We review the existing literature from three perspectives: the benefits of joint text--chart representation of data, the existing tool support for synthesis of such representation, and the use of natural language processing in visualization systems to coordinate text and charts.

\subsection{Symbiotic Relationship of Text and Visualization}

The synergetic association of text and visualization is essential for telling compelling stories with data~\cite{bach2017emerging,riche2018data}. Traditional analytical visualizations focus on rapid analysis and exploration of data.%such as identifying trends or outliers and supporting domain experts.
They leverage perceptually effective channels to encode data, but usually do not involve textual elements. On the other hand, explanatory visualizations emphasize conveying messages to a broader audience---text plays a crucial role. Existing research indicates that people tend to allocate significant attention to the text while using these visualizations~\cite{borkin2015beyond}. 

Several existing studies have investigated the benefits of combining text and visualizations. Zhi et al.~\cite{zhi2019linking} studied the impact of visualization--text linking---visual marks in a chart are highlighted when hovering over a relevant text phrase. In a controlled experiment, they found that participants recall information better when it is interactively linked across both representations, although they did not observe improvement in comprehension. In contrast, Barral et al.~\cite{lalle2019gaze} evaluated a gaze-driven approach---rather than explicit linking~\cite{zhi2019linking}---and found that it improves comprehension for low-literacy participants~\cite{barral2020understanding}. %Similarly, \textit{Elastic Documents}~\cite{badam2018elastic} is an interactive viewer augmenting text and tables with on-demand contextual visualizations and, when compared to a conventional PDF viewer, it was found that it improves the quality of summarization as well as comprehension to a moderate extent. 
When studying the impact of explicit visualization--text linking in the context of a Bayesian reasoning problem, Ottley et al.~\cite{ottley2019curious} found that people tend not to consolidate information well across the text and visualizations, suggesting the need for a better support of content integration. 

Based on this empirical evidence, our aim is to technically support the creation of interactive references between text and charts. Unlike the specific examples of previous studies, we investigate full-fledged authoring support with automatic suggestions in an interactive document editor.

\subsection{Systems for Authoring Explanatory Visualizations}

For communicating complex data, data-driven storytelling---which combines text and visualization into engaging visual data stories---has been advocated~\cite{lee2015more,segel2010narrative,kosara2013storytelling}. Unlike analytical tools that focus on rapidly generating data visualizations, tools that help create such data stories allow users to add textual descriptions and annotations, as well as to customize visual marks and layouts~\cite{stolte2002polaris,wongsuphasawat2015voyager,mackinlay1986automating,cook2005illuminating}. Some of these tools focus on the production of a single narrative visualization~\cite{satyanarayan2014lyra,ren2014ivisdesigner,liu2018data,ren2018charticulator,xia2018dataink,datawrapper,kim2016data,kim2019dataselfie} or addition of direct annotations~\cite{ren2017chartaccent}. Others support creating a complete narrative with a sequence of visualizations and textual explanations~\cite{gratzl2016visual,amini2016authoring,satyanarayan2014authoring,brehmer2019timeline,kim2019datatoon, mcneill2019viz}. Existing systems also explore novel forms of data stories including videos~\cite{amini2016authoring}, comics~\cite{kim2019datatoon}, and slideshows~\cite{gratzl2016visual,satyanarayan2014authoring,brehmer2019timeline}. 

However, existing storytelling tools do not support the interactive synthesis of text and visualization. In these tools, textual parts are mostly considered passive supportive elements---semantically connected yet separated from the associated visualizations. Therefore, practitioners rely on programming frameworks (e.g., D3~\cite{bostock2011d3} or Idyll~\cite{conlen2018idyll}) in order to create interactive references between the two (see an interactive article about Boston's subway system\footnote{\url{http://mbtaviz.github.io/}}). But this requires advanced programming skills. Research prototypes showcase the benefit of visualization--text integration using specific examples \cite{kwon2014visjockey,lalle2019gaze,badam2018elastic,steinberger2011context}, but do not support the authoring process. Scientific publishers such as \textit{Authorea}~\cite{authorea} and \textit{Elsevier}~\cite{elsevier} attempt to support this integration, but are limited to very simple linking (e.g., finding a related figure given an explicit figure reference in text). Kong et al.~\cite{kong2014extracting} recently used crowdsourcing to reconstruct references between textual phrases and visual marks on the charts in existing documents. Latif et al.~\cite{latif2018exploring,latif2019authoring} suggest a programmatic, but declarative syntax using HTML attributes to establish interactive references.

In contrast to existing tools and programming frameworks, we target users who do not have programming expertise and aim to provide an accessible user interface for creating interactive links between text and charts. A recent work~\cite{sultanum2021leveraging}, published while our work was being reviewed, provides a comparable interface, but the linking method is purely image-based (e.g., color-based region selection). Our approach leverages underlying data semantics to enable a more expressive linking strategy with intelligent data-driven assistance.

% To force this figure to appear in the page 3
\begin{figure*}[tb]
 \centering
 \includegraphics[width=1\textwidth]{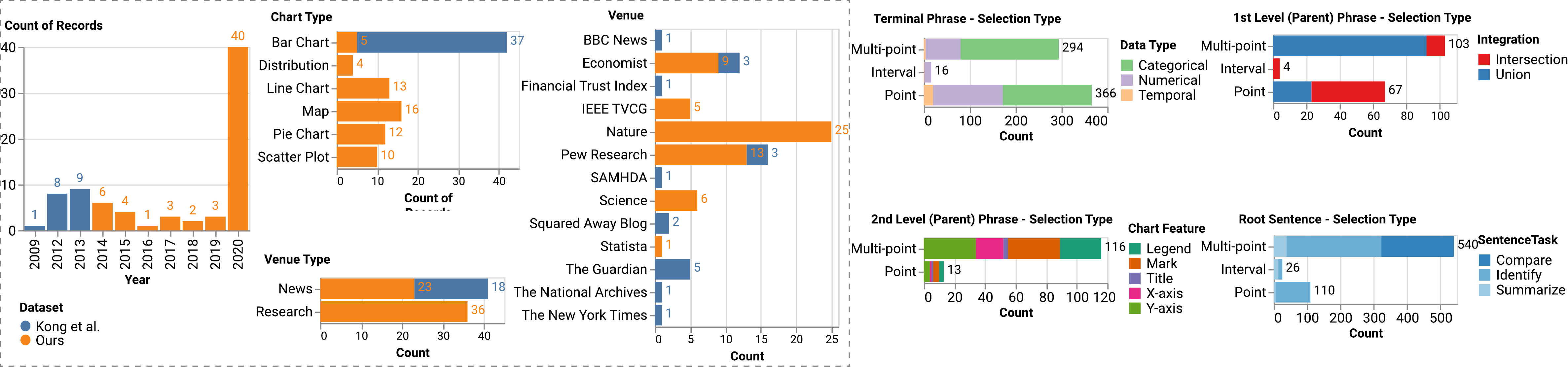}
 \caption{Overview of our data collection. We augmented existing articles from Kong et al.~\cite{kong2014extracting} with additional venues and chart types (left). We observed point, multi-point, and interval selections in the text--chart references (right). They are often grouped together to generate aggregate selections, while also forming a hierarchical reference tree: \textit{terminal} phrase $\rightarrow$ \textit{1\textsuperscript{st} level} parental phrase $\rightarrow$ \textit{2\textsuperscript{nd} level} parental phrase $\rightarrow$ \textit{root} sentence; see \autoref{fig:ref-example} for an example.}
 \label{fig:data-collection}

\end{figure*}

\subsection{Natural Language Understanding in Visualization}

Our work leverages natural language processing to automatically infer potential references between text and charts. 
%Natural language understanding is still under-explored in visualization systems and only recently, we see an influx of research leveraging it to tackle specific subproblems. 
Natural language processing has already been used to interact with visualizations, for instance, for assisting data analysis and exploration tasks~\cite{srinivasan2017natural} or question-answering about statistical charts~\cite{kim2020answering}. Only a few systems use it for establishing referencing between text and charts. Among these, \textit{Elastic Documents}~\cite{badam2018elastic} extract words from a document and perform simple keyword matching to create interactive links to generated charts; Kim et al.~\cite{kim2018facilitating} offer a similar method for tables. Similarly, Lai et al.~\cite{lai2020automatic} automatically annotate raster images of bar, pie, and scatter plots using user-generated text. Provided an entity name and its characteristics, they identify the relevant visual area using keyword matching.

In contrast, we focus on establishing references between existing text and charts in a \textit{communication} context. Although \textit{Elastic Documents}~\cite{badam2018elastic} and Lai et al.'s work~\cite{lai2020automatic} tackle a similar problem, they are limited to template-based matching for point-level references and a small set of charts. We aim to support a variety of charts and address interval-based references in addition to point references. Our system also generates higher-order groupings of the point and interval references following the syntactic parsing of a sentence (\autoref{fig:ref-example}).

\input{design_space}

\section{Design Considerations}
\label{sec:design}

Based on the findings of our design space analysis, we designed \toolname{}, an authoring tool, with a focus on establishing explicit referencing between a chart and related text. The main objective is to facilitate non-technical users in creating customizable and highly interactive data-driven documents. The system should assist users by detecting and suggesting potential references while they are composing a document. Besides, the system should provide an intuitive and easy-to-use interface for the manual construction of references beyond the automatic suggestions. We developed \toolname{} along the following design rationale.

\bpstart{D1. Suggest possibilities} The system should not explicitly create references, rather suggest them to the user. One reason is that every automatic detection system would inevitably result in false positives as the matter is complex and the linking might be subjective to the user. Another reason is that users may get annoyed by the automatic creation of many references---not all wanted. Therefore, the system should only suggest them, and the user can either accept or reject them. 

\bpstart{D2. Let users create} Users should not feel restricted to only what is suggested by the system---the automatic suggestion of potential references may not be enough. Users may want to create additional references or combine the suggested ones into a single higher-level reference. The construction of a reference involves the selection of visual marks (data items) on the chart and relating them to the text. To this end, our goal is to support smooth visual interactions for creating references. The interaction design should be intuitive, efficient, and generalizable to multiple chart types.

\bpstart{D3. Assist but do not distract} In general, the users should not get distracted by the additional features the tool provides. Users should be able to focus on creating the content while assistance for linking text and charts blends in smoothly. It might go unnoticed first but become a valuable tool when revising and polishing the document. The suggestions and options to create references might even inspire the authors to communicate a deeper analysis of the data as an understandable communication is easier to achieve.

\section{\toolname{} Usage}
\label{sec:usage}
\toolname{} comprises an \textit{editor} and a \textit{viewer}. We discern two roles of users: \emph{authors}, who create the content within the editor, and \emph{readers}, who consume the content in the viewer. While \toolname{} assists authors in creating interactive references, the readers profit from an improved synthesis of text and charts leveraging those interactive references. In the following, we describe a usage scenario that illustrates both perspectives, before \autoref{sec:system} discusses the technical details. For the sake of demonstration, we take an example dataset and a collection of charts that describe the Covid-19 cases in the US (on federal and state levels) from January to late August 2020. Let us assume that Alice writes an article for Bob.

\bpstart{Authoring} \autoref{fig:teaser} shows the editor interface of \toolname{}. It consists of a chart gallery and an editing area. The chart gallery  holds a collection of charts that can be inserted into the editor \circleBlue{1}. %\toolname{} supports charts based on Vega-Lite syntax~\cite{satyanarayan2017vega-lite}. Charts can be loaded into the gallery by dragging-and-dropping their specification files. Alice begins her composition by adding four charts \circleBlue{1} she has created using \textit{Voyager}~\cite{wongsuphasawat2015voyager}---a \textit{Tableau}-like interface for data analysis that can export Vega-Lite charts. 
The editor window provides a standard text formatting toolbar at the top. Charts can be dragged from the gallery to the editor. 

Alice first wants to give an overview of the temporal development and adds the line chart (state level) to the editor \circleBlue{2} and then starts writing text. While typing, she gets automatic suggestions (e.g., New York). Special decorations---dotted gray underline---notify her about the suggestions. Curious, she hovers over the suggestion ``New York'' to preview it. As a result, the line representing New York gets highlighted \circleBlue{2}. Unsure about the spelling of Illinois, she types the `@' symbol to trigger available suggestions while creating a reference for Illinois \circleBlue{3}. The small chart avatars in the suggestion panel notify her about what chart the suggestion corresponds to. Alice observes an interesting pattern about cases dipping toward the end of April and wants to create a reference for that. She does so by first selecting the text phrase and activating the reference construction mode \circleBlue{4}. Then, she selects the line chart in the \textit{link setting} panel \circleBlue{5}, chooses the filtering mode~\inlinegraphic{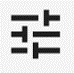}, adjusts the date value of the interval slider, and finalizes her reference construction.

\begin{figure*}[tb]
 \centering
 \includegraphics[width=0.98\linewidth, trim=0.15in 1.25in 3.4in 0.15in, clip=true]{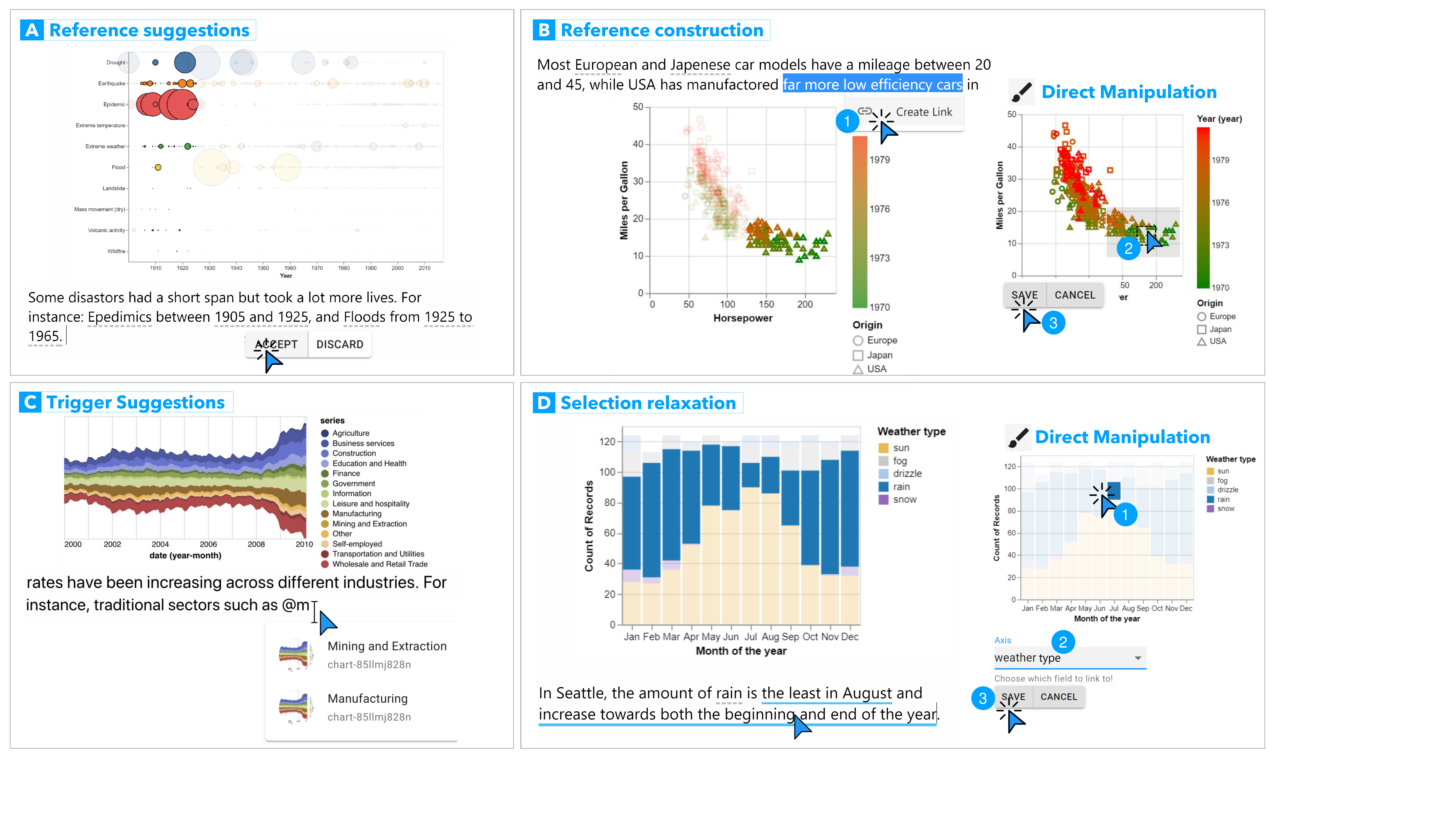}
 \caption{Salient features of Kori. (\textbf{A}) It uses natural language processing to suggest potential references between charts and the text while a user types. (\textbf{B}) Users can construct references by directly manipulating~\inlinegraphic{brush} the visual marks on the chart. (\textbf{C}) It is possible to manually trigger suggestions. (\textbf{D}) In direct manipulation~\inlinegraphic{brush} mode, users can easily expand their current selection to multiple visual marks of the same type. The encircled numbers mark the sequence of interactions for each activity.}
 \label{fig:kori-features}
\end{figure*}

Moving forward, she wants to provide a comparison of the three states that were hit hard by the pandemic during the second wave. She sees a suggestion for each state name, but she wants to highlight them at once to allow comparison. She simply does so by following similar steps as \circleBlue{4} and \circleBlue{5} but this time choosing the direct manipulation~\inlinegraphic{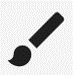} mode (see \autoref{fig:kori-features}\rectBlue{B}), and simultaneously selecting three corresponding timelines using a multi-point selection on the chart \circleBlue{7}. Since these references are manually constructed, they are underlined with blue color.

To give an impression of the severity of the pandemic, she adds a map of total deaths to the editor \circleBlue{8}. As she describes the states suffering more than 10,000 causalities, she gets a suggestion \circleBlue{9}. She previews it to see whether it is correct and then accepts it. As she previews it, she realizes the opacity of faded-out regions was quite low, and they were hardly visible. She then adjusts the inactive opacity as desired \circleBlue{\small{10}}.

\bpstart{Reading} Bob reads the composed article in the viewer. It is a restrictive version of the editor and offers a reading interface where Bob can activate the explicit referencing by interacting with the reference text. We use visual highlighting to make the related parts of a chart stand tall. The default highlight scheme is the opacity channel.

\section{The \toolname{} System}
\label{sec:system}

%The implementation of \toolname{} includes the challenges of coming up with a mixed-initiative interaction paradigm that assists human authors (D1) and does so in a passive and non-distracting manner (D3) while still giving full control and freedom to the authors (D2). 

\toolname{} is a web-based system. The front-end is developed in JavaScript, React, Draftjs, and Vega-Lite, while the back-end is implemented in Python using Flask. The natural language processing tasks have been performed in Python using SpaCy and FastText~\cite{mikolov2018advances}.

 \begin{figure*}[tbh]
 \centering
 \includegraphics[width=\textwidth]{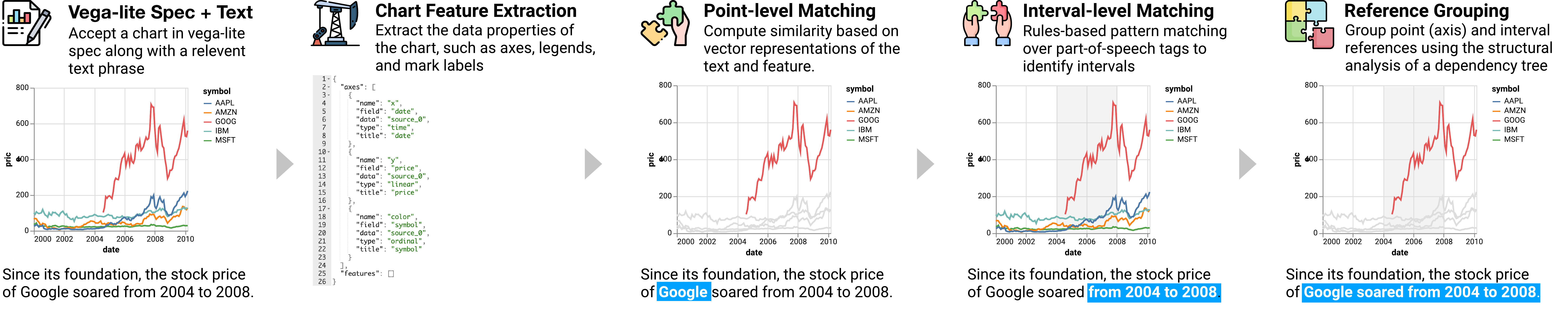}
 \caption{Four stages of our automatic reference suggestion pipeline. It accepts text and Vega-Lite specifications as input and begins with extracting features of a chart. These features are then matched against user text to find point references. The third step identifies numerical intervals in the user text. Finally, related references are grouped together to form higher level references.}
 \label{fig:auto-linking}
\end{figure*}

% Icons made by \textit{Freepik} from \url{www.flaticon.com}

\subsection{Charts}
We support a wide range of chart types, including but not limited to (stack or group) bar charts, (multi) line charts, scatter plots, distribution plots, heat maps, and choropleth maps. However, advanced visualizations (e.g., network diagrams, treemaps, etc.), as well as charts with coordinated views, are not supported. Similarly, interactive charts are not supported as existing interactions might conflict with our reference construction mechanism. %\toolname{} expects Vega-Lite~\cite{satyanarayan2017vega-lite} specification---JSON syntax---of a chart. 
We rely on Vega-Lite~\cite{satyanarayan2017vega-lite} for constructing and modifying charts as it offers an expressive and declarative syntax. Authors can load their data into Voyager~\cite{wongsuphasawat2015voyager} or Vega-Lite editor~\cite{vegaEdit} to construct and export charts. These charts can then be imported to \toolname{} by dragging the specification files and dropping them into the gallery. %Although not implemented, the chart gallery can be connected to \textit{Voyager}\cite{wongsuphasawat2015voyager} to further facilitate data analysis and chart construction.

\subsection{Reference Detection}

We define text--chart reference as an explicit linking of a text phrase to corresponding visual marks. Since both representations correspond to the same data items, our text--chart links are instantiations of conceptual cross-referencing between the charts and text. \toolname{} supports \textit{point}, \textit{multi-point}, and \textit{interval} references.

An automatic reference detection intends to speed up the composition process. We detect all three types of references and group references by combining numeric intervals with the corresponding axes of the chart. Once the references are identified, their presence is communicated to the author via visual cues without interrupting the authoring process (D1, D3); they are underlined with a dotted gray line. Authors can inspect them in due time or safely ignore them as they would not appear in the viewer mode. Authors can preview them by hovering over before accepting or discarding (\autoref{fig:kori-features}\rectBlue{A}). Once accepted, they are shown with blue underline and will appear in the viewer as interactive references. \autoref{fig:auto-linking} shows our automatic reference detection approach that consists of the following steps:

\bpstart{Chart Feature Extraction} Charts are composed of different types of encoding that map data values to visual properties (e.g., position, color, shape) of marks. The first step is to extract these data properties (e.g., axes labels, axes values and scales, legends, data labels) of a chart. The expressive JSON syntax of Vega-Lite enables easy access to this information. We loop through all kinds of visual encodings and extract their values in the underlying data. We also extract the axes properties as this is particularly relevant for interval references. The extracted features serve as a knowledge base to match the user-typed text against and for suggesting potential references. For instance, the extracted features of the scatterplot in \autoref{fig:kori-features}\rectBlue{B} are: x and y axes (\textit{horsepower} and \textit{miles per gallon}), legend categories (Europe, Japan, and USA), and name of all car models (denoted by each dot).

\bpstart{Point-level Matching} The next step detects the occurrences of chart features in a user-typed text. Our matching process uses vector representations of text obtained from FastText~\cite{mikolov2018advances}, a neural network-based approach to obtain text representations. It takes into account both word- and subword-level information (and therefore more resistant to noise than word-only approaches such as word2vec~\cite{NIPS2013_9aa42b31}). In addition to exact keyword matches, this vector-based matching process can tackle typographical errors, slight variations of the words or phrases (e.g., US, U.S., USA), abbreviations (e.g., EU, European Commission), synonyms (e.g., donating, contributing), and semantically similar words (e.g., Obama, Democrat).
We first use FastText to obtain vector representations of (i) chart features and (ii) $n$-gram representations in the sentence up to $n=5$. For each chart feature, we then select the $n$-gram in the sentence that had the highest cosine similarity to the chart feature and present it as a potential link to the user if the threshold is greater than 0.5. Both the $n$-gram size and the similarity threshold were selected empirically to maximize the F$_1$ score against a small set of 55 manually-annotated examples (see \autoref{fig:nlp-eval}).

\bpstart{Interval-level Matching} A rules-based approach processing the words and part-of-speech tags is used to identify intervals in a sentence. We derived heuristics based on our observations in the design space analysis (\autoref{sec:design-space-analysis}). We observed 116 sentences (16 from our collection, 100 gathered from diverse news articles on the web) that described one or more numerical intervals each and found the following list of frequent patterns (their frequencies in brackets):

\noindent \texttt{more/less/fewer than X (45), X to/through Y (29), between X and Y (15), at least X (11), since/from X (4), below/above X (3)}

\noindent The symbols \texttt{X} and \texttt{Y} denote either a number, date, or time, which is identified using the parts-of-speech tag \texttt{NUM} from Spacy. The combination of words and part-of-speech tags makes it convenient to derive compact rules to capture intervals. For instance, the rule \texttt{`between X and Y'} is identified with the part-of-speech pattern \texttt{`NUM CCONJ NUM'}, which identifies phrases where two numbers (\texttt{NUM}) are joined by a coordinating conjunction (\texttt{CCONJ}); a complete list of patterns is contained in the supplemental material.

\begin{figure*}[tb]
 \centering
 \includegraphics[width=0.98\textwidth]{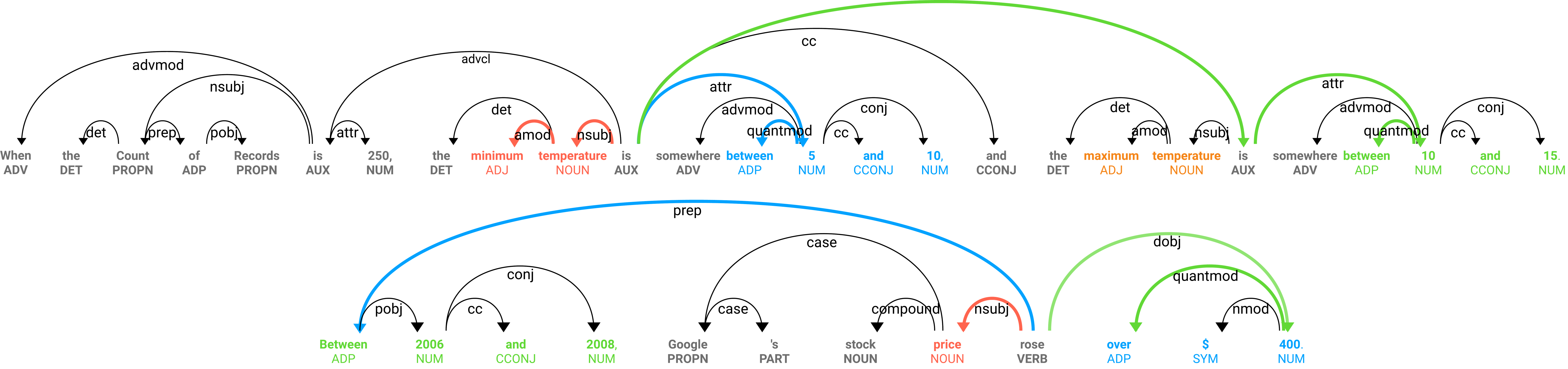}
 \caption{Dependency parsing of two sample sentences. Success case: \texttt{minimum temperature} is successfully grouped with \texttt{between 6 and 10} (top). Failure case: \texttt{price} is wrongly grouped with \texttt{between 2006 and 2008} as it closer to \texttt{price} than \texttt{over \$400} (bottom).}
 \label{fig:dep-tree}
  \vspace{-\intextsep}
\end{figure*}

\bpstart{Reference Grouping} The final step is to combine the interval with the correct axis-reference on the chart. In some cases, we can infer this by comparing the interval occurrence to chart axis values, for example, in simple charts with a single numerical axis. However, if a sentence contains multiple intervals or a chart has multiple axes with overlapping numeric scales, inferring the correct interval--axis combination is not trivial. Surface-level heuristics that, for example, combine an interval with the nearest axis reference are often inadequate as they do not take into account the structure of the sentence. To better account for sentence structure, we use its dependency tree (again obtained from Spacy)
to map interval occurrence to their corresponding axis references. Concretely, we map an interval occurrence to the axis reference that is closest in the dependency tree distance, where we treat the dependency tree as an undirected graph and use Dijkstra's algorithm to compute the distance between words. For cases where axis reference and/or intervals consist of multiple words,  we compute the tree distance between the phrase head words. As an example, in \autoref{fig:dep-tree} (top), the sentence contains two axis names ($A_1=$ \texttt{minimum temperature}, and $A_2=$ \texttt{maximum Temperature}) of a scatter plot and two intervals ($I_1=$ \texttt{between 5 and 10} and $I_2=$ \texttt{between 10 and 15}). The distance between $A_1$ and $I_1$ is four (orange and blue arcs) and, between $A_1$ and $I_2$, it is five (orange and green arcs). Vice versa, the distance between $A_2$ and $I_1$ is five and, between $A_2$ and $I_2$ it is four. Combining an axis and interval with minimal distance, $A_1$ is paired with $I_1$ and $A_2$ with $I_2$, as desired. In case of ties, we use the distance to the first common ancestor as a tiebreaker (i.e., the interval--axis combination that shares a closer ancestor is grouped together). If this second heuristic results in a tie, too, then we resort to the surface-level distance---number of words between two candidate phrases---as the final tiebreaker. The approach works in many cases but has limitations---the bottom sentence in \autoref{fig:dep-tree} highlights a failure case where our approach wrongly groups the axis \texttt{price} with the interval \texttt{between 2006 and 2008} (distance 2 -- orange and blue arcs) instead of combining it with \texttt{over \$400} (distance 3 -- orange and green arcs).

\subsection{Reference Construction}

Often, authors already have references in mind while writing about a chart. Thus, they may not have to wait for reference suggestions. Instead, they can explicitly trigger suggestions by typing the `@' symbol (\autoref{fig:kori-features}\rectBlue{C}). It shows a list of features or labels of charts that are currently active in the editor. Each item in the suggestion list is preceded by a thumbnail of the related chart; this helps authors quickly see what feature refers to which chart. The list offers an auto-complete; it keeps on filtering as the user types. The list of suggestions is also triggered by selecting a portion of text. 

The reference suggestions are limited and may not cover the full spectrum of possibilities that an author needs. Authors may want to reference a certain part of the chart, a few distinct visual marks that seem interesting in some context, or combine arbitrary sets of visual marks related to a message they are communicating. \toolname{} offers a smooth interface to accomplish this using two distinct modes as shown in \autoref{fig:teaser}\circleBlue{4} -- \circleBlue{6} and \autoref{fig:kori-features}\rectBlue{B}.

The first is the \textit{direct manipulation} mode~\inlinegraphic{brush} (\autoref{fig:kori-features}\rectBlue{B}). Authors can select a portion of text \circleBlue{1} and then directly brush visual marks of interest \circleBlue{2}. The system enables only meaningful selections for a chart (e.g., no rectangular brushing for maps). \toolname{} also offers point and multi-point selections. Having selected a set of visual marks, authors can finalize the reference construction \circleBlue{3}. The selection of visual marks is enhanced through relaxation of a selection~\cite{heer2008generalized}. Relaxing a selection is a way of saying \textit{``select all items like this one''}. This is particularly useful in situations where users want to reference many visual marks that are of the same type as a single or couple of selected marks. \autoref{fig:kori-features}\rectBlue{D} explains an example where an author likes to create a reference to data dimension \texttt{weather type = rain}; she can simply do so by selecting any blue rectangle (\texttt{rain}) \circleBlue{1} and relaxing the selection to \texttt{weather type} by selecting it in the \textit{Axis} (data dimension) drop-down \circleBlue{2}.

While the direct manipulation mode is natural, it is not that flexible, especially for selecting overlapping marks or precisely combining multiple data dimensions (shown in a chart). Complementing this, the \textit{filtering} mode ~\inlinegraphic{selection} (\autoref{fig:teaser}\circleBlue{4} -- \circleBlue{6}) allows 
setting a filter for every data dimension used in a chart. Values of the dimensions can be adjusted using a multi-selection search field for categorical variables or an interval slider for numerical ones. A combination of multiple filters corresponds here to the creation of a higher-level reference using  \textit{intersection} operations (see also \autoref{sec:design-space-analysis}).

\toolname{} uses opacity as the default visual highlighting scheme when previewing the interactive references as opacity does not mostly interfere with the existing visual encoding. Each chart is provided with a configuration panel---like the one in \autoref{fig:teaser}\circleBlue{10}---where the degree of opacity can be modified. Also, users can choose the fill highlighting; we provide a color picker to select colors for active and inactive marks.

\section{Evaluation}
To evaluate our approach, we tested the automatic reference detection pipeline and conducted a user study.

\begin{figure}[tb]
     \centering
     \includegraphics[width=\columnwidth]{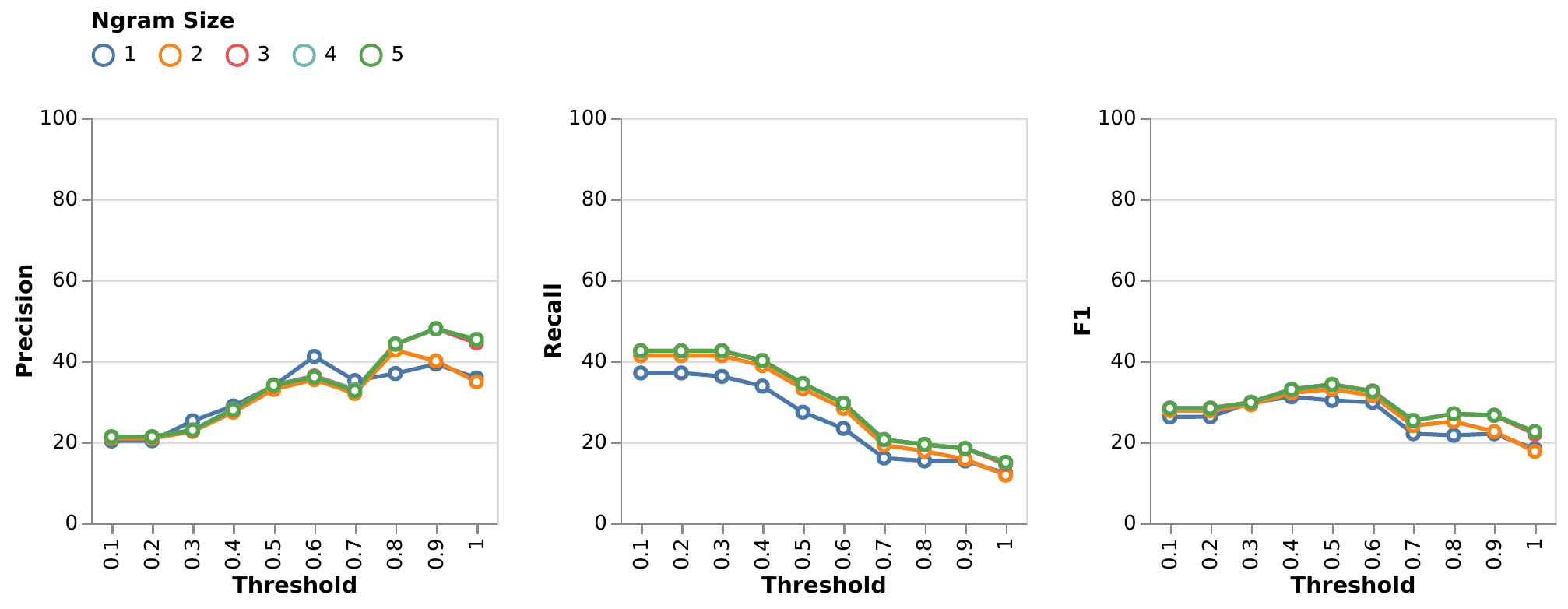} % 3 in a row
     \caption{Quantitative evaluation of the reference detection approach. Precision, recall, and $F_1$ at various maximum $n$-gram sizes for the validation dataset (55 text--chart pairs).}
     \label{fig:nlp-eval}
      \vspace{-\intextsep}
\end{figure}

\subsection{Algorithmic Evaluation: Reference Detection}
To quantitatively evaluate our reference detection approach, we needed pairs of Vega-Lite chart specifications and corresponding text. Although we had collected 110 text--chart pairs (\autoref{sec:design-space-analysis}), we did not have access to the underlying data for charts in all those examples. Therefore, we curated a separate dataset for evaluation.

\subsubsection{Dataset Curation}
We had data for 42 of 110 text--chart pairs and their images (Kong et al.'s~\cite{kong2014extracting} collection). We re-constructed these charts in Vega-Lite. However, they were all bar charts, and included very few interval references. To increase the diversity and size of the sample, we added 50 pairs as follows: From our data collection (\autoref{sec:design-space-analysis}), we extracted 116 instances of sentences---including interval references---describing a variety of different charts. Additionally, we collected 34 diverse charts from example galleries of different visualization libraries (e.g., Vega, Vega-Lite, D3, Observables, etc.). Then, we mapped these charts to instances of the 116 sentences. One co-author manually rephrased the sentences to match the data on the chart yet keeping the essence as close to original sentences as possible. Another co-author, then, went through these sentences to make sure they are both syntactically and semantically correct. Finally, we annotated each text--chart pair for point, interval, and group references to create the ground truth. The resulting dataset includes 92 chart--text pairs with 15 different types of charts. This dataset was randomly split into a validation (60\%, 55 pairs) and a test (40\%, 37 pairs) dataset.

\subsubsection{Results}
\autoref{fig:nlp-eval} shows precision, recall, and $F_1$ score (harmonic mean of precision and recall) for varying similarity thresholds and maximum $n$-gram sizes on the validation dataset. We selected the threshold of 0.5 and the maximum $n$-gram size of 3 to maximize the $F_1$ score (0.34). %$F_1=1$ would mean that the system correctly suggested all references, while $F_1=0$ means no correct suggestion.
%Since too many incorrect references can bother the users, another option could be to go for higher precision ($n=5$, $threshold=0.8$). This results in fewer (32) false positives.

%% Results on Test dataset
%Ngram Size / Threshold / Accuracy / Precision / Recall / F1 
%3	0.5	22.8	37.5	36.77	37.13
%% At n=3, threshold = 0.5; maximized F1
%TP/FP/FN 57/90/93 Total (150)

%% Results on Kong et al. Dataset
%Ngram Size/Threshold/Accuracy/Precision/Recall/F1       
%3       0.5     27.13   34.96   54.78   42.68  

For the test dataset, our pipeline correctly identified 57 references (out of 137 true references), produced 90 incorrect references (false positives), and missed 93 references (false negatives). 
\toolname{} correctly suggested 42 of 84 point, 9 of 26 interval, and 5 of 27 group references. While our approach worked better for identifying point references ($F_1=0.47$), the interval detection ($F_1=0.25$) and reference grouping ($F_1=0.26$) were more challenging.

When running our approach on Kong et al.'s dataset (42 text--chart pairs), we obtained an average distance ($1-F_1$; lower values are better) of 0.57 from the ground truth (annotated examples by experts) compared to 0.39 produced by their approach. Although quantitatively, references extracted by Kong et al. are closer to those in the ground truth, we detect references automatically while they rely on user intervention to extract base references (distance to the ground truth of 0.54) and then automatically refine them that reduces the distance to 0.39.

\subsubsection{Discussion}

Several problems contribute to the lower $F_1$ scores in our approach. One problem with general purpose pre-trained vector representations of text is that it matches words that are too dissimilar in the context of a chart (e.g., \textit{``Obama''} matches to both chart categories \textit{``Democrats''} and \textit{``Republicans''} with cosine similarity of 0.58 and 0.56 respectively). A similar problem occurs with numbers. Numbers are important in dealing with charts and can be described in different ways (e.g., 12, twelve). FastText had problems matching \textit{``60''} to \textit{``sixty''} in the sentence \textit{``Most movies have a rating between sixty and 100''}, and instead matched \textit{``60''} to \textit{``100''}, presumably because Arabic numeral representations of numbers are closer in the vector space. While parts-of-speech tagger of Spacy \texttt{NUM} could identify numbers with units, it has no support for dates (e.g., months, days); they are tagged as proper nouns (\texttt{NNP}). Therefore, intervals like \texttt{Apr to Jun} could not be detected. Some failure cases in a grouping, like the one in \autoref{fig:dep-tree} (bottom sentence), can be avoided if we consider the extents, scales, and units of numerical axes in the chart in addition to distances in the dependency tree. 

% For point references: F1 = 0.47, For interval: F1 = 0.25

\begin{figure*}[tb]
 \centering
 \includegraphics[width=\linewidth, trim=0.9in 1.8in 0in 0in, clip=true]{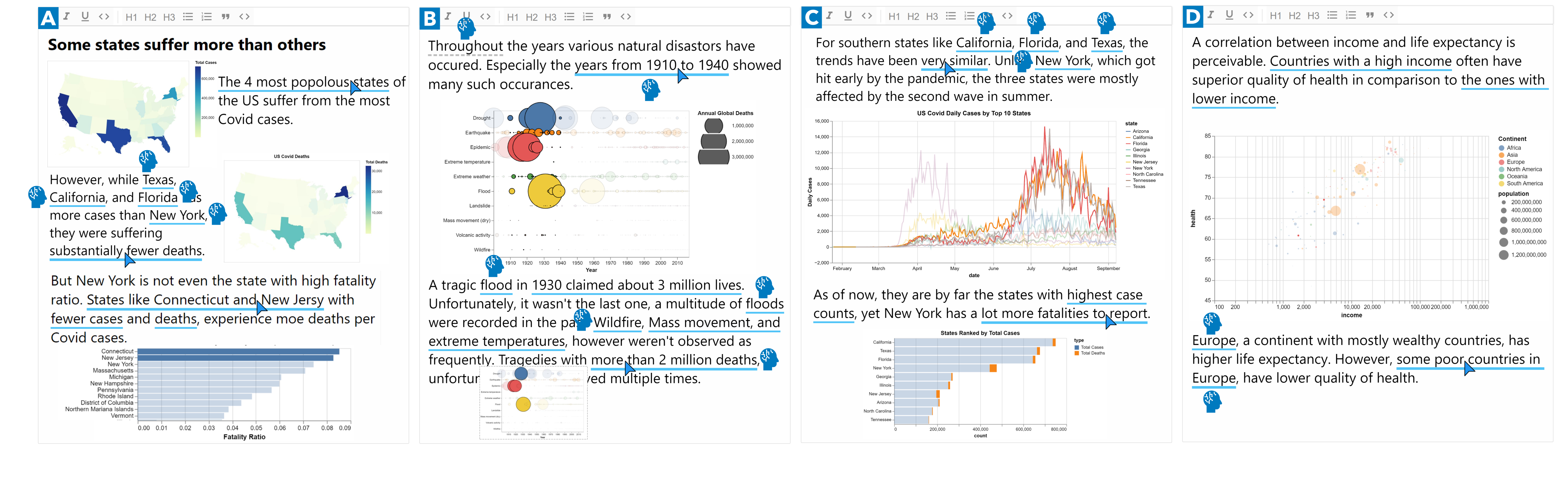}
 \caption{Copy-edited excerpts of stories created by participants. The symbol~\inlinegraphic{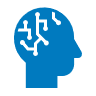} marks the reference suggestions. (\textbf{A}) P3 relied on reference construction using direct manipulation~\inlinegraphic{brush}. (\textbf{B}) In contrast, P8 got 9 suggestions (1 wrong). She manually constructed 2 references: \textit{``1930 claimed about 3 million lives.''} and combined two suggestions (\textit{``mass movement''} and \textit{``extreme temperatures''}) into a single reference. (\textbf{C}) P5 used first reference \textit{``very similar''} to verify a fact he was describing. (\textbf{D}) P6 got only two suggestions and created most references using filtering~\inlinegraphic{selection} interface.}
 \label{fig:user-stories}
 \vspace{-\intextsep}
\end{figure*}

\subsection{User Study: Authoring}

We conducted a user study to gain insights into the usefulness and usability of our system. We focus on evaluating how people interacted with the system to author an interactive document. We could not follow a comparative evaluation approach because comparative systems (e.g., work of Sultanum et al.~\cite{sultanum2021leveraging}) were not available.
%While most existing works do not support or evaluate such an authoring process~\cite{badam2018elastic,kong2014extracting}, future work can compare our system with the recently published similar work by Sultanum et al.~\cite{sultanum2021leveraging}.

\subsubsection{Participants}
We recruited 11 participants (P1--P11; five male and six female) with diverse backgrounds ranging from undergraduate students (3 -- P2, P9, P10) and graduate students (3 -- P5, P6, P11) to visualization experts (4 -- P1, P3, P4, P8) and a user interface designer (P7). All participants had experience using document editing tools like Word, Google Docs, or similar. All participants mentioned that they had created charts using data science toolkits (e.g., R, Python) or programming libraries (e.g., D3, Vega-Lite). All participants except P10 regularly created charts as part of their job or studies. Eight participants mentioned to have worked with charts in a word processing tool (e.g., Google Doc, Word). 

\subsubsection{Procedure and Tasks}
In sessions lasting about 60 minutes each, the participants used Google Chrome on their personal computers to access the tool in an online video call with an experimenter, sharing their screens. After collecting the above demographic information, every session began with a brief introduction of the project followed by a short tutorial. We demonstrated the main features of \toolname{} using a variety of chart types. 

In the main part of the study, the participants had to complete three tasks. First, to familiarize themselves with the system, we asked them to replicate an example from the tutorial, which contained a bar chart and a paragraph of text with three interactive references. Second, the participants had to reproduce a given but previously unseen example. Two charts (a scatterplot and a heatmap) and two paragraphs of text were provided with nine references. We marked the references in the text, and participants had to transform them into interactive references according to their understanding. We tried to maximize the diversity of references so that participants had to use different features of \toolname{} to construct them. In the third task, the participants had to design a short story (5 to 6 sentences) on one of the three scenarios (they were free to pick any) that were provided to them in the form of one or more charts.

We concluded the session with a reflection survey on the usability and usefulness, rating different statements on a 5-point Likert scale (1 -- strongly disagree, 5 -- strongly agree). Moreover, we discussed the overall user experience, potential improvements, and limitations in semi-structured interviews. 

\subsubsection{Results}

All participants successfully completed all three tasks within the limited session time.
They created references in Task 1 (3 references) and Task 2 (9 references) with minimal intervention from the experimenter. For these two tasks, \toolname{} suggested 6 references (2 for Task 1 and 4 for Task 2) as expected. In Task 3, every participant created a short story with one or more charts; \autoref{fig:user-stories} shows four examples. In total, \toolname{} suggested 64 references~\inlineboxplot{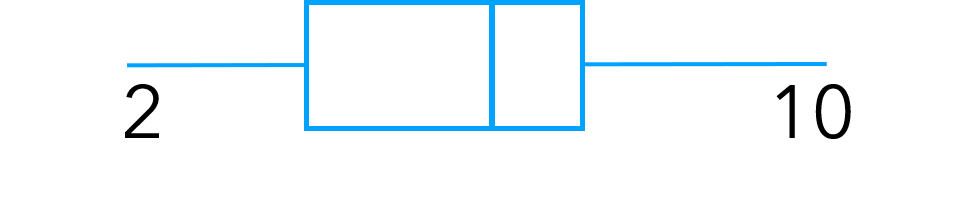} for Task 3. (The small boxplots show the distribution of suggestions for all 11 participants). These 64 suggestions also include 25 instances where participants explicitly triggered suggestions. Among these, 48 references (including 25 explicitly triggered suggestions, all correct point references) were correct~\inlineboxplot{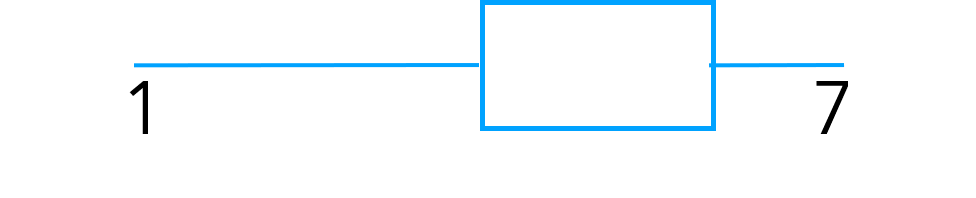} and 16 incorrect~\inlineboxplot{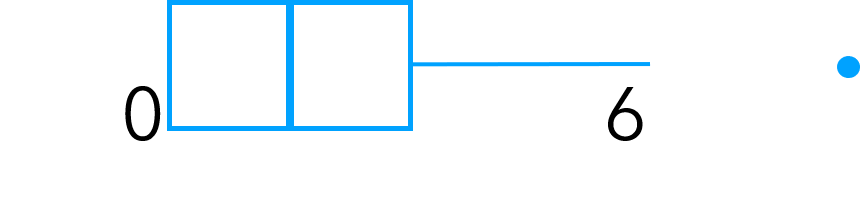}. The incorrectly suggested references were rarely ignored (3) and mostly discarded (13). Besides, the participants manually created 40 references~\inlineboxplot{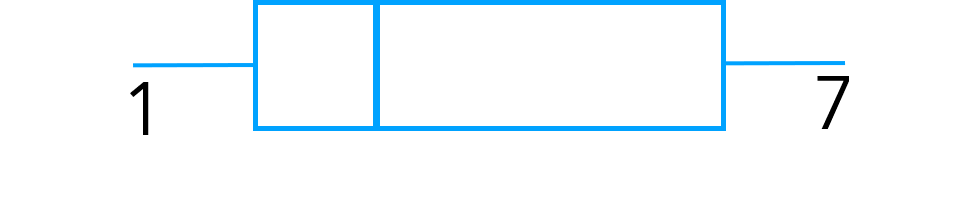}.

For point references, they often relied on automatic suggestions. We observed comparatively fewer instances of explicit triggering by typing `@', but more selecting a text phrase. While participants P3 (\autoref{fig:user-stories}\rectBlue{A}) and P6 (\autoref{fig:user-stories}\rectBlue{D}) largely relied on manual construction of references, P8 (\autoref{fig:user-stories}\rectBlue{B}) made use of more automatic suggestions.
We observe the frequent use of direct manipulation mode~\inlinegraphic{brush} for geographical maps, bar charts, and line plots. Comparatively, the participants employed filtering mode~\inlinegraphic{selection} more frequently for scatter plots and bubble charts. 

\bpstart{Overall experience} Participants highly rated the overall experience (median, mode = 4; IQR = 0; \inlinegraphic{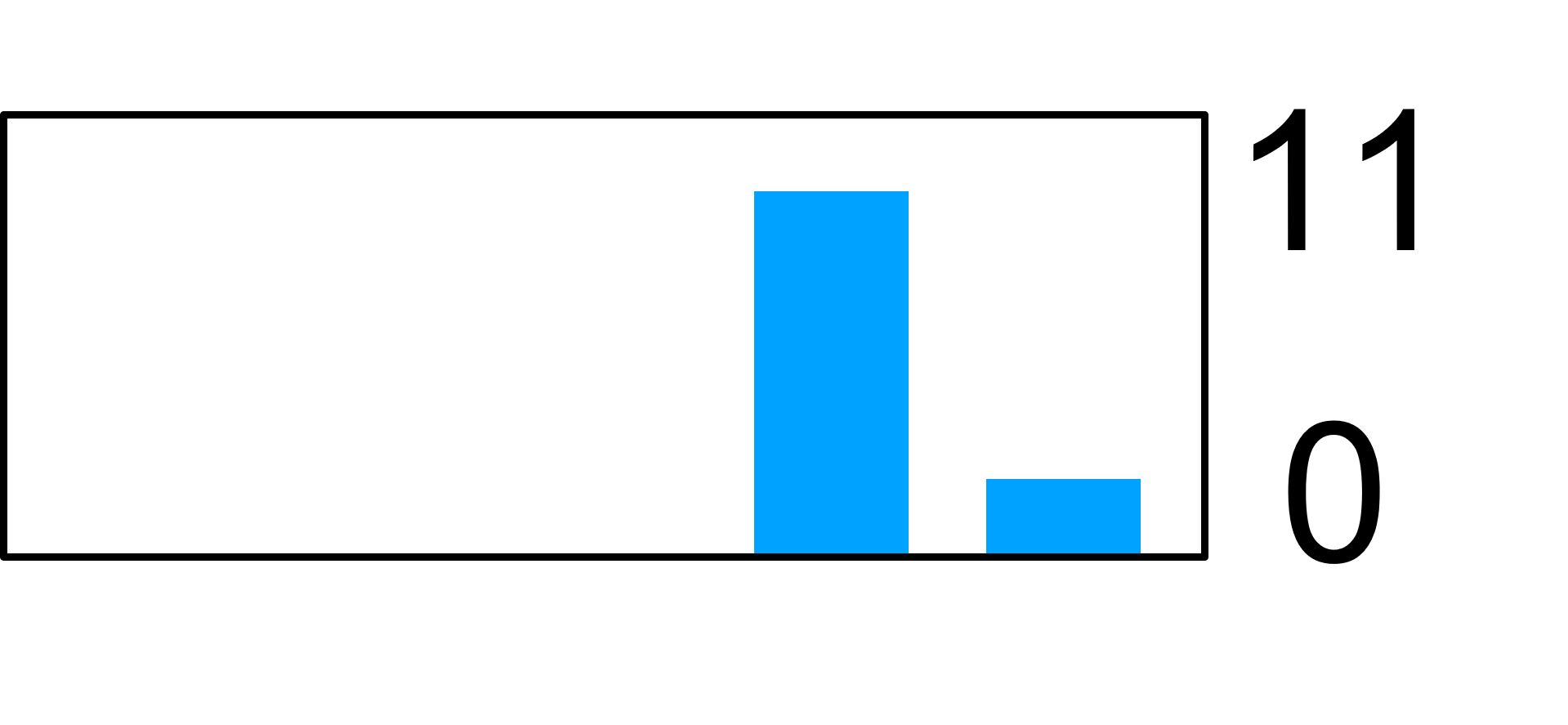}) and usefulness of the tool (median, mode = 5; IQR = 0; \inlinegraphic{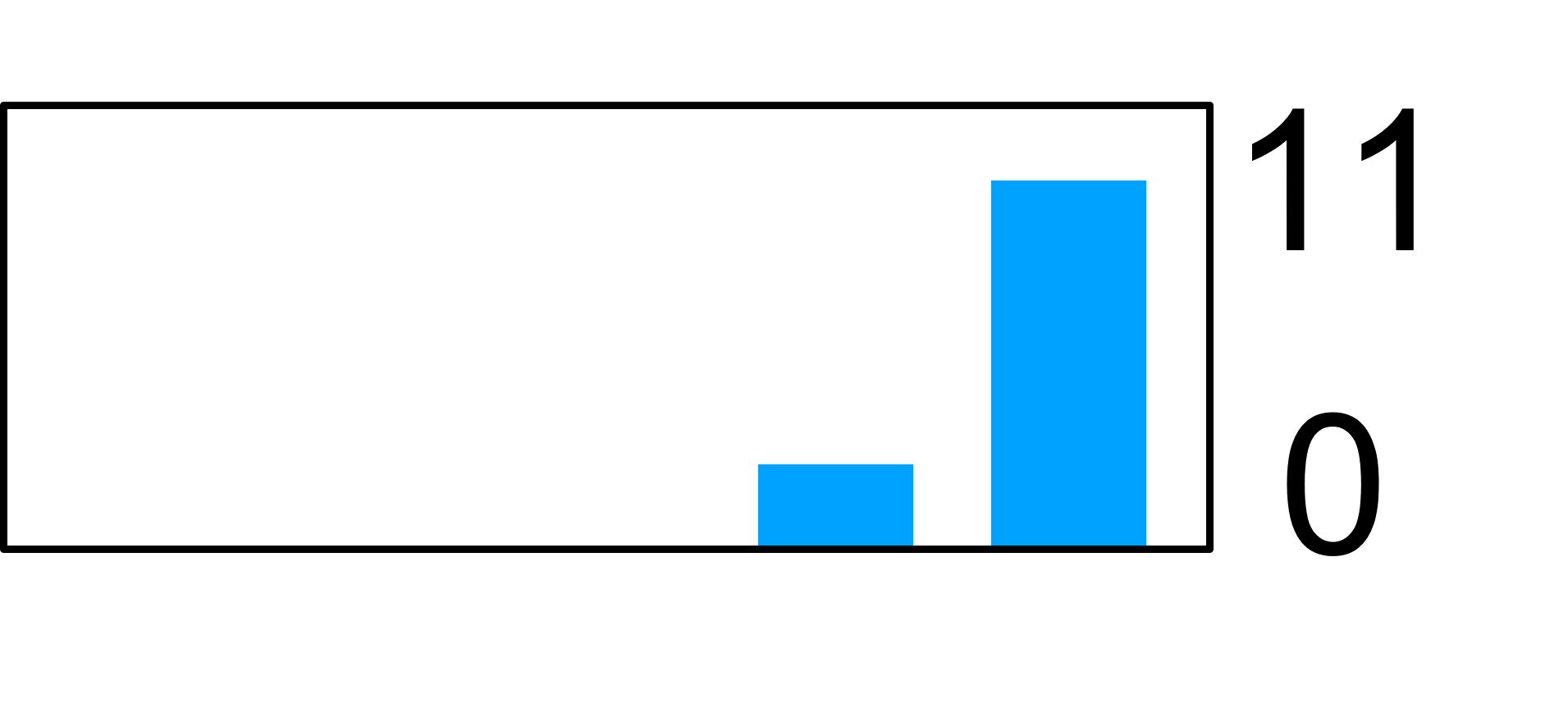}). While participants found all features of the tool intuitive and self-explanatory, they rated the reference construction interface (median, mode = 4; IQR = 1; \inlinegraphic{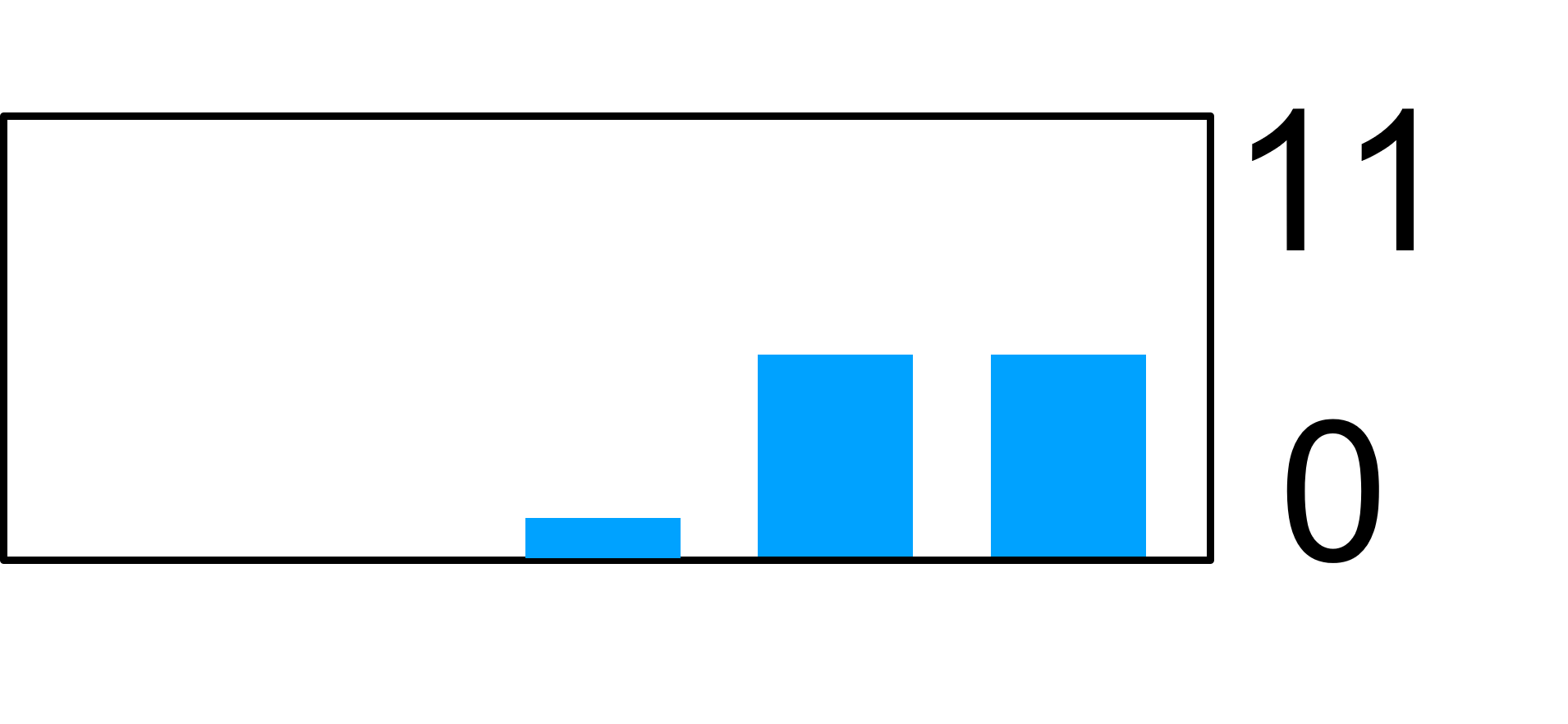}) higher than the automatic suggestions (median, mode = 4; IQR = 0; \inlinegraphic{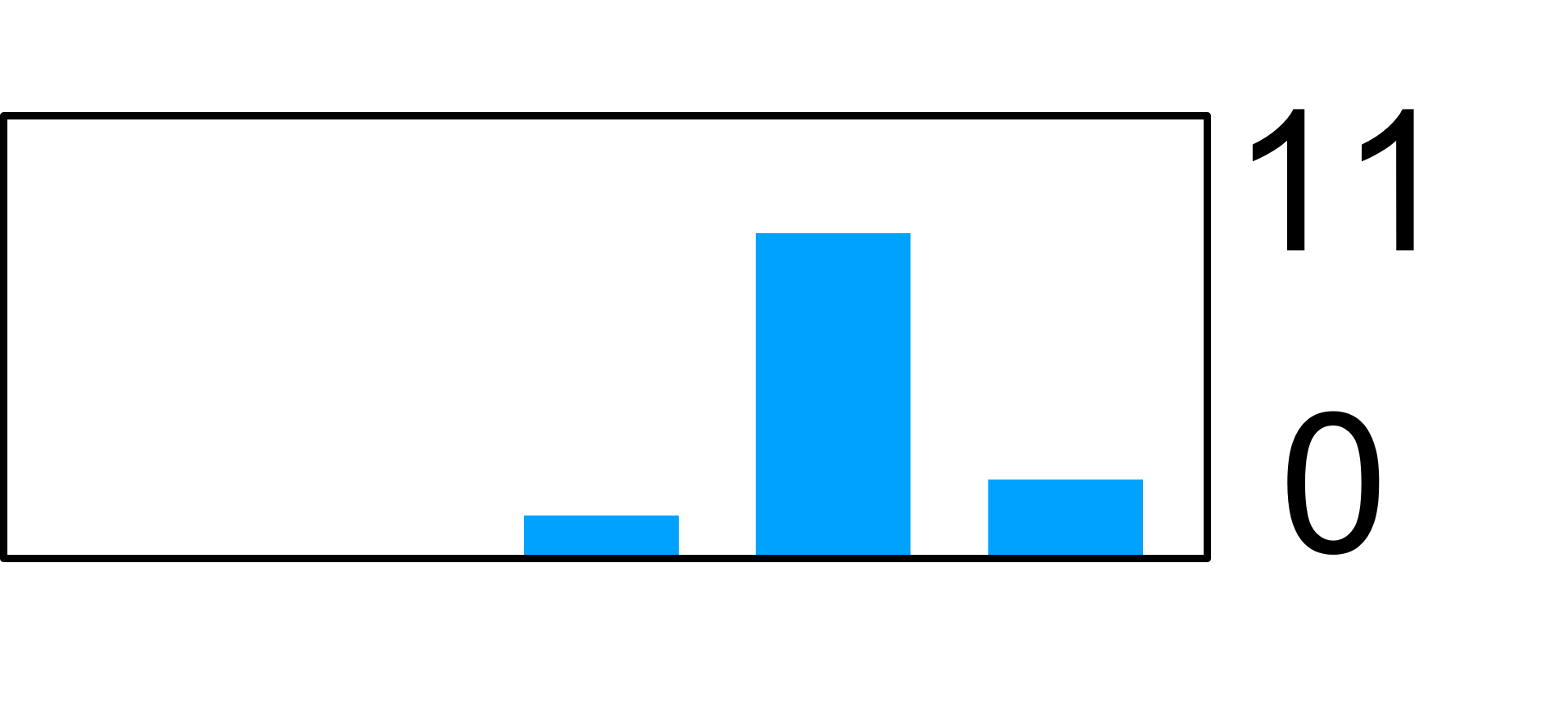}).

\bpstart{Manual and Automatic Linking}
All participants mentioned that the tool is intuitive, easy to use, and has high learnability. Two participants (P3, P4) even said they would have easily discovered all the features without a demonstration. Participants specifically liked the reference construction interface. P4 praised the direct manipulation~\inlinegraphic{brush} of regions on the map to construct a reference and described it as easy as \textit{``click regions of interest and done. Impressive!''}. P5 stated \textit{``It is surprising that manual linking is so flexible and works with so many chart types.''} P6 appreciated the two different modes of reference construction. P11 favored the filtering~\inlinegraphic{selection} mode over direct manipulation, saying it is more flexible and precise. The participant further elaborated: \textit{``Finding one data point in millions of data points is just impossible in the brushing interface.''} P9 complained about not being able to brush the color legend of a chart (\toolname{} only supports the direct manipulation of marks inside the chart area). 

All participants agreed that the automatic suggestions are valuable and complement the reference construction yet criticizing that they are not `smart' enough. P9 stated that automatic suggestions are helpful and work reliably for simple cases. P1 commented \textit{``suggestions are cool but not too smart.''} Nonetheless, none of the participants considered suggestions as distracting. On the contrary, P2 and P7 suggested making their presence more noticeable. However, wrongly identified references were a bother for one participant (P11). Three participants (P2, P7, P10) complained about delays of automatic suggestions. P8 remarked that the suggestions are often too short, and their grouping needs to be better supported using the interface---the participant wanted to group two references that were correctly suggested in a sentence (\toolname{} does not support grouping of suggested references).

\bpstart{Surprises and Future Directions}
Surprisingly, several participants (P5, P6, P9, P11) came up with a different use case that we did not have in mind. They used interactive references as a means to explore the data. P5 described his experience as \textit{``I wasn’t sure about a statement''}---Task 3: he had an assumption that California, Florida, and Texas have similar trends for Covid-19 (\autoref{fig:user-stories}\rectBlue{C})---\textit{``and I could verify it using interactive linking. Readers can do too.''} P6 reported that interactive linking provides insights and helps in understanding the data. She described her opinion as \textit{``I would say it’s not only about writing but also [for] doing analysis while writing.''}. Two participants (P1, P5) suggested having an embedded chart creation interface. Without such interface, it is limited to what could be explored, P1 pointed out and said, \textit{``Charts were limited and non-modifiable. [Kind of] limits the scope of investigative reporting.''}

Several participants highlighted the lack of some convenience features, such as editing a reference that has been suggested or created, deleting a reference without deleting its text, and a user interface to group the suggested references.

\bpstart{Usefulness} 
Most participants (P1--P6, P11) saw \toolname{}'s value in creating information for reporting, presentation, and communication scenarios. P11 highlighted the benefit of interactive references as \textit{``Readers may not interpret my intention correctly. This tool makes it much clearer by creating explicit references to make sure readers look at what I intended.''} Similarly, P8 mentioned that visual highlighting through interactive links may be better than adding direct annotations on top of a chart, which might increase visual clutter.

\section{Discussion}

Reflecting on the results of the evaluation, we discuss the limitations of the current solution and directions for future research. This also leads us to a broader discussion on the interplay of exploration and explanation, as well as externalizing cognition through visualization in the context of data-driven storytelling.

\subsection{Limitations and Opportunities}
%The user study also revealed several limitations of our current tool that provide future research opportunities. 
Currently, our automatic suggestion does not resolve the user's intention on which chart to link to. As a result, we observed that it often does not link to the chart expected by the user, especially when the charts share the same underlying dataset. Resolving such ambiguity can be challenging without the user's explicit input. A potential resolution method would consider the user's current cursor or the distance from the text to the chart. A related issue is that we currently only support one-to-one mapping, resulting in a unidirectional link from one text phrase to a single chart. However, it is unclear whether supporting one-to-many or many-to-many mapping models is desirable. Not only the authoring interface can become more complicated, but also the benefit for reading may be limited or even adversarial. A user study could shed light on this.

Our user study focused on the authoring interface as it is our core contribution compared to existing research focusing on the reading experience. However, once the links are established, we foresee many opportunities to provide an improved reading experience beyond simply revealing references upon hovering a text phrase. For instance, the user might want to see all relevant text phrases linked to a specific chart for a better understanding of context. To meet this need, the tool can annotate related visual marks in the chart with the relevant text phrases. This is similar to the \textit{gather} operation in exploring embedded word-scale visualizations by Goffin et al.~\cite{goffin2020interaction}, while it is the opposite approach to \textit{Elastic Documents}~\cite{badam2018elastic} collecting relevant visualizations for a selected phrase.

A major technical challenge is to make the automatic linking as reliable and accurate as possible. Due to the small size of our dataset, we relied on various heuristics to identify the text--chart links. While some of these already leverage recent advances in natural language processing (e.g., a state-of-the-art neural dependency parser trained over representations from pre-trained language models~\cite{devlin-etal-2019-bert}), others are simpler (e.g., template-based interval pattern matching). Our approach does not learn a joint model that integrates all these heuristics into one. Therefore, it would be interesting to explore a more data-driven approach that trains an end-to-end model from a body of text and its associated charts directly from the raw input. Such a system would require collecting a substantial training set of annotated examples, but may allow for the use of more sophisticated contextualized embedding models (e.g., BERT or Transformers as opposed to static FastText) that can be subsequently fine-tuned on the collected training set.

Integrating further automation is another promising direction. For instance, the generation of text that describes the data is possible~\cite{srinivasan2018augmenting,latif2019vis}, and there has also been a flurry of recent work on data-driven table-to-text generation with deep learning techniques~\cite{wiseman-etal-2017-challenges}. Adapting such methods, we could suggest prompts about the interesting data facts in the context of chart-to-text generation.

\subsection{Closing the Loop from Explanation to Exploration}
One interesting insight we learned from the user study is that participants often use the authoring interface to explore data. \toolname{} currently assumes that the users already explored the data and prepared charts for drafting the story. Based on this assumption, \toolname{} does not support chart creation and thus provided pre-made charts to study participants. 
Since, in the simplified setting of the study, participants did not work with the data before, they used automatically suggested references as opportunities to examine details of the data and also actively inspected the data by trying out different selections in the manual construction interface. This behavior was particularly frequent for high data density in charts without data labels (e.g., a scatterplot with a multitude of data points). Our observation demonstrates a potential need for augmenting \toolname{} to support data exploration, supporting the full lifecycle of data-driven storytelling. An interesting question is how we can support creating additional relevant charts starting from text phrases or existing references. That means, from explanation, going back to exploration and coming forth again. 

\subsection{Improving External Cognition in Data-driven Articles}

Visualizations play a role as an external cognitive aid, offloading the mental load of memorizing and processing data into an external visual representation. Nowadays, visualizations are embedded into diverse media, bringing additional challenges for reading, from efficiently allocating our attention to synthesizing related information across different modes. How we can reduce the cognitive burden incurred from multi-modal information is an exciting avenue for future research. In this work, we address the split-attention effect between text and charts primarily through interactive highlighting. According to the cognitive load theory and the cognitive theory of multimedia learning, our highlighting approach follows the signaling principle or temporal contiguity principle~\cite{mayer201412}. Both theories suggest alternative approaches. For instance, instead of integrating information at the temporal dimension (i.e., through highlighting on demand), we can leverage the spatial dimension as well, reducing the spatial proximity between the text and charts (e.g., using word-sized graphics integrated into the text~\cite{beck2017word}). Incorporating additional information modalities such as audio comments~\cite{Latif2021talking} might also be helpful to further reduce the distance in the information space.

\section{Conclusion}
We developed \toolname{}, a mixed-initiative system to support authors in creating references between the text and charts. We analyzed text--chart references in existing articles to inform the design of \toolname{}. Findings on point and interval references, as well as the grouping of the references, guided the development of the automatic suggestion pipeline. A flexible manual interface provides a complementary way to construct references. An algorithmic evaluation and a user study demonstrate the benefits of \toolname{} for easily creating references with respect to different types of datasets and visualizations, as well as revealed limitations. Future work includes improving the reliability and accuracy of our automatic suggestion pipeline, while also improving the capability of the reference construction interface beyond the unidirectional linking from the text to a chart. Further evaluation of \toolname{} in different system configurations may quantitatively analyze the interplay of manual and automated linking methods. Translating our approach to other forms of data stories such as data comics~\cite{kim2019datatoon} provides relevant future directions.

\acknowledgments{
We wish to thank the reviewers for their constructive feedback. The project is funded by the Deutsche Forschungsgemeinschaft (DFG, German Research Foundation) – 424960846.}

\bibliographystyle{abbrv-doi-hyperref}

\bibliography{paper}
\end{document}

%% file: design_space.tex
\section{Understanding Text--Chart References}
\label{sec:design-space-analysis}

To understand how text and charts are referenced in data-driven documents, we analyzed a collection of news articles and scientific papers that contain text--chart pairs.

\subsection{Data Collection and Analysis}

We started with an initial collection of documents from Kong et al.~\cite{kong2014extracting}. It was limited to bar charts alone from 18 articles gathered from different news media. We expanded this collection with additional chart types and sources, resulting in 77 articles with 110 paragraph--chart pairs. We initially targeted three main sources: visualization papers, research articles published in Nature, and news articles. Within these sources, we manually picked examples for maximizing the diversity of charts. \autoref{fig:data-collection} shows the distribution of examples in our collection.

We divided each paragraph--chart pair into sentence--chart pairs for our analysis, resulting in 227 pairs. Following the same process as Kong et al.~\cite{kong2014extracting}, we manually constructed \textit{minimal} references between text and charts. A reference is minimal if it cannot be subdivided into meaningful smaller references. Two researchers followed an open coding process to analyze the sentence--chart pairs. The researchers used the collection of Kong et al.~\cite{kong2014extracting} as a basis to derive the initial codes. These codes were then applied to our collection. The resulting collection is an extended version that contains additional codes and greater diversity of publication venues and chart types. %, and we iteratively resolved any conflicts that arose during the process to reach a consensus.

\subsection{Analysis Results}

We report the results along five semantic groups of findings, building on each other. The focus of our analysis is always how text, data, and visualization interact with each other.

\begin{figure}[tb]
 \centering
 \includegraphics[width=1.0\linewidth]{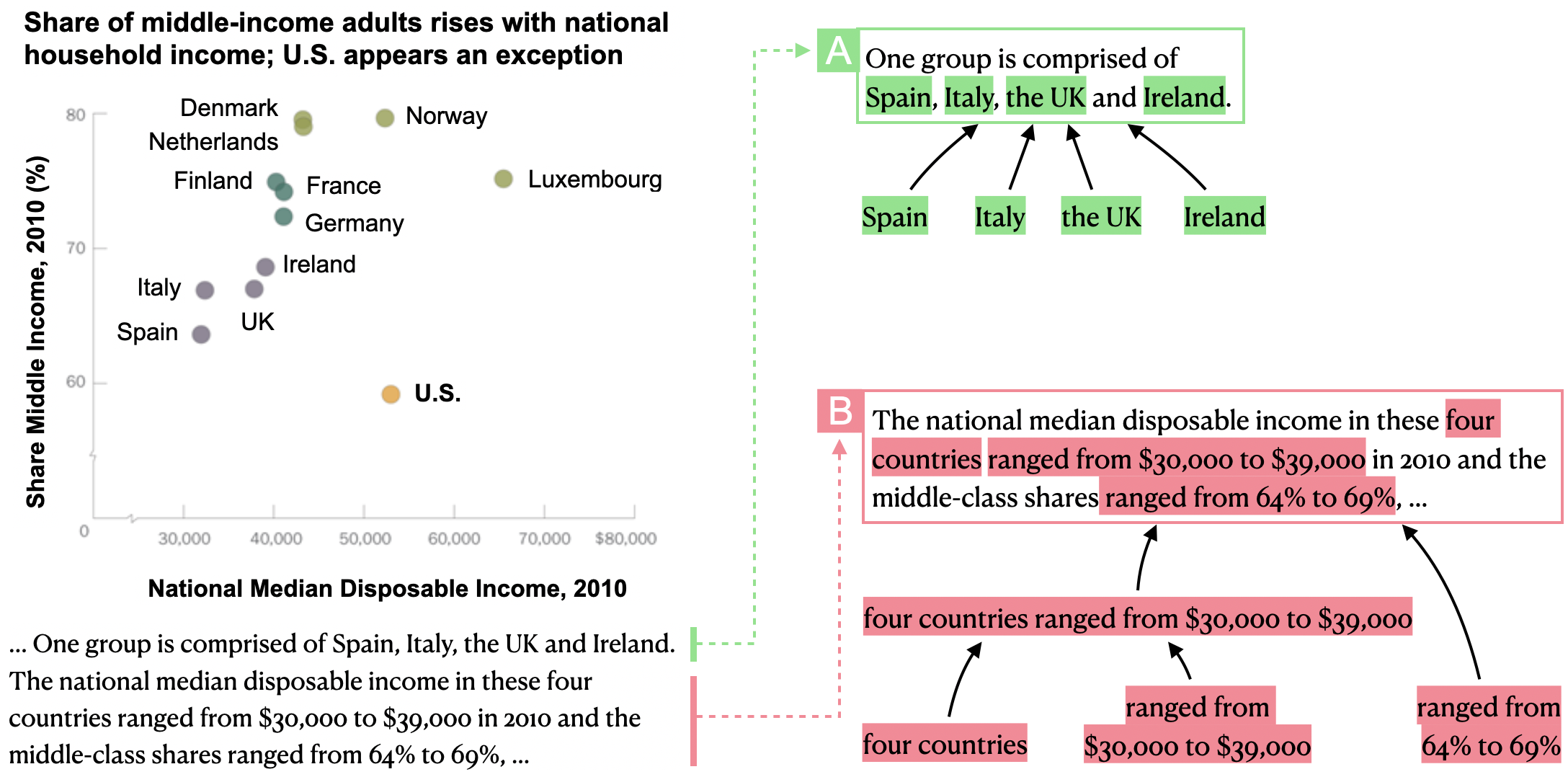}
 \caption{Excerpt of an article from Pew Research~\cite{examplearticle}. Demonstration of text--chart reference grouping: The colored text phrases are minimal references, while \textbf{A} and \textbf{B} show how they can be grouped hierarchically.}
 \label{fig:ref-example}
 \vspace{-\intextsep}
\end{figure}

\bpstart{References resemble selections} A key insight is that every text--chart reference bears a similarity to a selection operation in a visualization. Just as a visualization offers \textit{point} selections (such as clicking one or more visual marks) and \textit{interval} selections (such as brushing a region of visual marks~\cite{satyanarayan2016vega}), a text reference can describe one or more visual marks by directly referencing item names or an interval of data points by mentioning scale extents. For instance, in \autoref{fig:ref-example}A, each country name in the phrase \textit{``Spain, Italy, the UK, and Ireland''} selects each corresponding mark in the scatter plot, while the phrase \textit{``ranged from \$30,000 to \$39,000''} in \autoref{fig:ref-example}B refers to all marks falling into the range.

Out of 676 identified minimal references, we observed 366 point, 294 multi-point, and 16 interval references. The referenced chart features mainly included axes (283), individual visual marks (262), and legends (109); references to axes and legends eventually lead to selecting a set of visual marks in the chart area. While the underlying data values of the referenced chart features are mostly categorical (407), numerical (244) and temporal (25) data values were also referenced. %The main trend holds for point and multi-point references---194 categorical, 152 numerical and 20 temporal in point references, and 213 categorical, 77 numerical and 4 temporal in multi-point references. 
The sixteen interval references relate to 15 numerical and one temporal data attribute. Interestingly, we observed only four (out of 676) \textit{visual} references such as ``red arrow'' or an ``orange slice''. This is maybe because visual encoding in standard charts is rather self-explanatory.

\bpstart{References are grouped hierarchically} We found that minimal references can be grouped together to create higher-order references. The grouping is similar to how a parse tree represents the syntactic structure of a sentence. In a parse tree, constituents---a word or a phrase---are terminal nodes that serve as independent units, e.g., verbs, nouns. A minimal reference is a terminal text phrase in our reference tree that establishes an independent connection to a chart feature. For instance, in \autoref{fig:ref-example}A, each country name is a minimal reference and can be combined with others to create a parent reference of the four countries. Eventually, the whole sentence lies at the root, referring to all the countries (\autoref{fig:ref-example}A). On the other hand, the sentence in \autoref{fig:ref-example}B has three minimal references: \textit{``four countries''}, \textit{``ranged from \$30,000 to \$39,000''}, and \textit{``ranged from 64\% to 69\%''}. The first two can be integrated to create a parent or non-terminal reference node. This parent node can then be combined with the third minimal reference \textit{``ranged from 64\% to 69\%''}, resulting in the root sentence (\autoref{fig:ref-example}B). When references are combined, the leftover text---which is not part of any minimal reference---should be added to the parent reference, which otherwise becomes fragmented. That is, the final sentence contains all text rather than only the text that represents minimal references. 

In our collection, we observed up to four levels of reference groupings, with the highest-level as root reference being the sentence itself. The minimal references lie at level 0 (leave nodes in the parse tree). There were 175 first-level and 28 second-level ancestor references, with an average of 2.48 and 2.68 minimal references, respectively. While 131 sentences had at least first-level references, only 27 (out of 227) sentences had second-level references. At non-terminal levels, we observed more multi-point references compared to single-point references---103 versus 67 at the first level and 25 versus six at the second level. We observed four interval references at the first level and no interval reference at the second level; this might partly be because we coded it as \textit{point} or \textit{multi-point} when an \textit{interval} reference was combined with either of them.

\bpstart{Reference grouping incurs data transformations} The grouping of multiple references results in either \textit{union} or \textit{intersection} of individual selections of data points. For instance, when the phrases \textit{``ranged from \$30,000 to \$39,000''} and \textit{``ranged from 64\% to 69\%''} referring to two different variables are combined, the resulting reference relates to the intersection of data points identified by the two ranges. On the other hand, grouping of \textit{``Spain''}, \textit{``Italy''}, \textit{``UK''},  and \textit{``Ireland''} results in a union of the corresponding data points. We observed more unions than intersections: 115 unions versus 59 intersections at the first level. The relative difference increases as we climb up the reference hierarchy: 27 versus four at the second level and 25 versus one at the sentence level.

\bpstart{References relate to visualization tasks} A text phrase in a reference can relate to standard visual analysis tasks~\cite{latif2021deeper}. All minimal references represent \textit{identification} tasks (e.g., identifying countries in \autoref{fig:ref-example}A), while representative visual marks of these tasks vary from a single-point (361) to multi-points (315). References at a higher level describe advanced tasks. Within the first-level references, we identified \textit{comparison} (33) and \textit{summarization} (12) tasks, in addition to \textit{identification} (130) tasks. On the second level, we observed \textit{comparison} (11) tasks, in addition to \textit{identification} (20) tasks. Overall, at the sentence level, we observed 149 \textit{identification}, 49 \textit{comparison}, and 29 \textit{summarization} tasks. %The tasks at a higher level typically involve more than a single point or multi-point references, e.g., describing an extreme data point, comparing multiple points, or summarizing a range of data points. 
An example of a \textit{comparison} sentence looks like \textit{``Some upper-middle-income countries, like the Dominican Republic and Thailand, seem to have deadlier roads than much poorer places such as Liberia''}~\cite{economist2020richest}, while an example \textit{summarization} sentence is \textit{``But when countries reach a GDP [...] of about \$30,000, death rates usually start to come down''}~\cite{economist2020richest}.

\bpstart{Ambiguities, transformations, and abstractions exist and often relate to data} As can be expected when dealing with natural language, we observed ambiguities in the text--chart references. Almost half of the reference phrases (311 of 676) exactly matched the labels in charts, and hence, were clear. Eighty-two text phrases were partial matches and ambiguous. For instance, the phrase \textit{``\$1 increase''} in a sentence partially matches to the chart label \textit{``assuming a \$1 increase in the minimum wage''}~\cite{economist2020higher}. Other reference variations include inferences (64)---e.g., \textit{``former communist states''} in text while the chart shows explicit state names. In our collection, the inference problem was more common among the references that were solely inferred from the visual encoding because the corresponding charts did not have textual data labels. We also observed potential ambiguity issues such as the use of synonyms (64), stemming/lemmatization (29), abbreviations (8), and hypernyms (1). For the phrases referring to numerical ranges, we frequently observed approximate numbers (21), especially for large numbers or numbers with decimal points. Another observation is that text phrases may also represent derived measures such as mean, variance, or other computed numbers (17)---e.g., \textit{``Nearly six-in-ten (58\%) in the U.S.''}~\cite{economist2020where}, which requires a transformation of the underlying data to be identified. Charts sometimes contain annotations showing these measures, but often they do not. %Therefore, it is another type of ambiguity that is equally challenging to resolve as other complex linguistic ambiguities.

\subsection{Limitations}

Although our collection is much more diverse and expansive compared to the initial dataset of Kong et al.~\cite{kong2014extracting}, the size of the sample and its representativeness are still limited, and it does not include advanced and coordinated multiple visualizations. Our qualitative analysis method could benefit from computational linguistic methods that may further discover semantic and structural insights into the space of text--chart references. We leave this for future work.

%% file: paper.bbl
\begin{thebibliography}{10}

\bibitem{authorea}
Authorea.
\newblock \url{https://www.authorea.com/}.
\newblock Accessed: 2020-08-10.

\bibitem{datawrapper}
Datawrapper.
\newblock \url{https://www.datawrapper.de/}.
\newblock Accessed: 2020-08-10.

\bibitem{vegaEdit}
Vega editor.
\newblock \url{https://vega.github.io/editor}.
\newblock Accessed: 2021-03-30.

\bibitem{amini2016authoring}
\href{https://doi.org/10.1109/TVCG.2016.2598647}{F.~Amini, N.~Henry~Riche,
  B.~Lee, A.~Monroy-Hernandez, and P.~Irani}.
\newblock \href{https://doi.org/10.1109/TVCG.2016.2598647}{Authoring
  data-driven videos with dataclips}.
\newblock \href{https://doi.org/10.1109/TVCG.2016.2598647}{{\em IEEE
  Transactions on Visualization and Computer Graphics}},
  \href{https://doi.org/10.1109/TVCG.2016.2598647}{23(1):501--510},
  \href{https://doi.org/10.1109/TVCG.2016.2598647}{2016}.
  \href{https://doi.org/10.1109/TVCG.2016.2598647}
{doi: {{%
10\hspace{.1pt}\discretionary{.}{%
}{.}\hspace{.4pt}1109\discretionary{/}{%
}{/}TVCG\hspace{.1pt}\discretionary{.}{%
}{.}\hspace{.4pt}2016\hspace{.1pt}\discretionary{.}{%
}{.}\hspace{.4pt}2598647}}}


\bibitem{ayres2012split}
P.~Ayres and G.~Cierniak.
\newblock Split-attention effect.
\newblock {\em Encyclopedia of the Sciences of Learning}, pp. 3172--3175, 2012.

\bibitem{ayres2005split}
P.~Ayres and J.~Sweller.
\newblock The split-attention principle in multimedia learning.
\newblock {\em The Cambridge Handbook of Multimedia Learning}, 2:135--146,
  2005.

\bibitem{bach2017emerging}
\href{https://doi.org/10.1109/MCG.2017.33}{B.~Bach, N.~Henry~Riche,
  S.~Carpendale, and H.~Pfister}.
\newblock \href{https://doi.org/10.1109/MCG.2017.33}{The emerging genre of data
  comics}.
\newblock \href{https://doi.org/10.1109/MCG.2017.33}{{\em IEEE Computer
  Graphics and Applications}},
  \href{https://doi.org/10.1109/MCG.2017.33}{37(3):6--13},
  \href{https://doi.org/10.1109/MCG.2017.33}{2017}.
  \href{https://doi.org/10.1109/MCG.2017.33}
{doi: {{%
10\hspace{.1pt}\discretionary{.}{%
}{.}\hspace{.4pt}1109\discretionary{/}{%
}{/}MCG\hspace{.1pt}\discretionary{.}{%
}{.}\hspace{.4pt}2017\hspace{.1pt}\discretionary{.}{%
}{.}\hspace{.4pt}33}}}


\bibitem{badam2018elastic}
\href{https://doi.org/10.1109/TVCG.2018.2865119}{S.~K. Badam, Z.~Liu, and
  N.~Elmqvist}.
\newblock \href{https://doi.org/10.1109/TVCG.2018.2865119}{{Elastic Documents}:
  Coupling text and tables through contextual visualizations for enhanced
  document reading}.
\newblock \href{https://doi.org/10.1109/TVCG.2018.2865119}{{\em IEEE
  Transactions on Visualization and Computer Graphics}},
  \href{https://doi.org/10.1109/TVCG.2018.2865119}{25(1):661--671},
  \href{https://doi.org/10.1109/TVCG.2018.2865119}{2018}.
  \href{https://doi.org/10.1109/TVCG.2018.2865119}
{doi: {{%
10\hspace{.1pt}\discretionary{.}{%
}{.}\hspace{.4pt}1109\discretionary{/}{%
}{/}TVCG\hspace{.1pt}\discretionary{.}{%
}{.}\hspace{.4pt}2018\hspace{.1pt}\discretionary{.}{%
}{.}\hspace{.4pt}2865119}}}


\bibitem{barral2020understanding}
\href{https://doi.org/10.1145/3377325.3377517}{O.~Barral, S.~Lall{\'e}, and
  C.~Conati}.
\newblock \href{https://doi.org/10.1145/3377325.3377517}{Understanding the
  effectiveness of adaptive guidance for narrative visualization: a gaze-based
  analysis}.
\newblock \href{https://doi.org/10.1145/3377325.3377517}{In {\em Proceedings of
  the 25th International Conference on Intelligent User Interfaces}},
  \href{https://doi.org/10.1145/3377325.3377517}{pp. 1--9},
  \href{https://doi.org/10.1145/3377325.3377517}{2020}.
  \href{https://doi.org/10.1145/3377325.3377517}
{doi: {{%
10\hspace{.1pt}\discretionary{.}{%
}{.}\hspace{.4pt}1145\discretionary{/}{%
}{/}3377325\hspace{.1pt}\discretionary{.}{%
}{.}\hspace{.4pt}3377517}}}


\bibitem{beck2017word}
\href{https://doi.org/10.1109/TVCG.2017.2674958}{F.~Beck and D.~Weiskopf}.
\newblock \href{https://doi.org/10.1109/TVCG.2017.2674958}{Word-sized graphics
  for scientific texts}.
\newblock \href{https://doi.org/10.1109/TVCG.2017.2674958}{{\em IEEE
  Transactions on Visualization and Computer Graphics (TVCG)}},
  \href{https://doi.org/10.1109/TVCG.2017.2674958}{23(6):1576--1587},
  \href{https://doi.org/10.1109/TVCG.2017.2674958}{2017}.
  \href{https://doi.org/10.1109/TVCG.2017.2674958}
{doi: {{%
10\hspace{.1pt}\discretionary{.}{%
}{.}\hspace{.4pt}1109\discretionary{/}{%
}{/}TVCG\hspace{.1pt}\discretionary{.}{%
}{.}\hspace{.4pt}2017\hspace{.1pt}\discretionary{.}{%
}{.}\hspace{.4pt}2674958}}}


\bibitem{borkin2015beyond}
\href{https://doi.org/10.1109/TVCG.2015.2467732}{M.~A. Borkin, Z.~Bylinskii,
  N.~W. Kim, C.~M. Bainbridge, C.~S. Yeh, D.~Borkin, H.~Pfister, and A.~Oliva}.
\newblock \href{https://doi.org/10.1109/TVCG.2015.2467732}{Beyond memorability:
  Visualization recognition and recall}.
\newblock \href{https://doi.org/10.1109/TVCG.2015.2467732}{{\em IEEE
  Transactions on Visualization and Computer Graphics}},
  \href{https://doi.org/10.1109/TVCG.2015.2467732}{22(1):519--528},
  \href{https://doi.org/10.1109/TVCG.2015.2467732}{2015}.
  \href{https://doi.org/10.1109/TVCG.2015.2467732}
{doi: {{%
10\hspace{.1pt}\discretionary{.}{%
}{.}\hspace{.4pt}1109\discretionary{/}{%
}{/}TVCG\hspace{.1pt}\discretionary{.}{%
}{.}\hspace{.4pt}2015\hspace{.1pt}\discretionary{.}{%
}{.}\hspace{.4pt}2467732}}}


\bibitem{bostock2011d3}
\href{https://doi.org/10.1109/TVCG.2011.185}{M.~{Bostock}, V.~{Ogievetsky}, and
  J.~{Heer}}.
\newblock \href{https://doi.org/10.1109/TVCG.2011.185}{D3 data-driven
  documents}.
\newblock \href{https://doi.org/10.1109/TVCG.2011.185}{{\em IEEE Transactions
  on Visualization and Computer Graphics}},
  \href{https://doi.org/10.1109/TVCG.2011.185}{17(12):2301--2309},
  \href{https://doi.org/10.1109/TVCG.2011.185}{Dec 2011}.
  \href{https://doi.org/10.1109/TVCG.2011.185}
{doi: {{%
10\hspace{.1pt}\discretionary{.}{%
}{.}\hspace{.4pt}1109\discretionary{/}{%
}{/}TVCG\hspace{.1pt}\discretionary{.}{%
}{.}\hspace{.4pt}2011\hspace{.1pt}\discretionary{.}{%
}{.}\hspace{.4pt}185}}}


\bibitem{brehmer2019timeline}
M.~Brehmer, B.~Lee, N.~H. Riche, D.~Tittsworth, K.~Lytvynets, D.~Edge, and
  C.~White.
\newblock Timeline storyteller: The design \& deployment of an interactive
  authoring tool for expressive timeline narratives.
\newblock In {\em Proceedings of the the Computation+ Journalism Symposium.},
  2019.

\bibitem{castro2019instructional}
J.~C. Castro-Alonso, P.~Ayres, and J.~Sweller.
\newblock Instructional visualizations, cognitive load theory, and visuospatial
  processing.
\newblock In {\em Visuospatial processing for education in health and natural
  sciences}, pp. 111--143. Springer, 2019.

\bibitem{conlen2018idyll}
\href{https://doi.org/10.1145/3242587.3242600}{M.~Conlen and J.~Heer}.
\newblock \href{https://doi.org/10.1145/3242587.3242600}{Idyll: A markup
  language for authoring and publishing interactive articles on the web}.
\newblock \href{https://doi.org/10.1145/3242587.3242600}{In {\em Proceedings of
  the 31st Annual ACM Symposium on User Interface Software and Technology}},
  \href{https://doi.org/10.1145/3242587.3242600}{UIST '18},
  \href{https://doi.org/10.1145/3242587.3242600}{pp. 977--989}.
  \href{https://doi.org/10.1145/3242587.3242600}{ACM},
  \href{https://doi.org/10.1145/3242587.3242600}{2018}.
  \href{https://doi.org/10.1145/3242587.3242600}
{doi: {{%
10\hspace{.1pt}\discretionary{.}{%
}{.}\hspace{.4pt}1145\discretionary{/}{%
}{/}3242587\hspace{.1pt}\discretionary{.}{%
}{.}\hspace{.4pt}3242600}}}


\bibitem{cook2005illuminating}
K.~A. Cook and J.~J. Thomas.
\newblock Illuminating the path: The research and development agenda for visual
  analytics.
\newblock Technical report, Pacific Northwest National Lab.(PNNL), Richland, WA
  (United States), 2005.

\bibitem{devlin-etal-2019-bert}
\href{https://doi.org/10.18653/v1/N19-1423}{J.~Devlin, M.-W. Chang, K.~Lee, and
  K.~Toutanova}.
\newblock \href{https://doi.org/10.18653/v1/N19-1423}{{BERT}: Pre-training of
  deep bidirectional transformers for language understanding}.
\newblock \href{https://doi.org/10.18653/v1/N19-1423}{In {\em Proceedings of
  the 2019 Conference of the North {A}merican Chapter of the Association for
  Computational Linguistics: Human Language Technologies, Volume 1}},
  \href{https://doi.org/10.18653/v1/N19-1423}{pp. 4171--4186}.
  \href{https://doi.org/10.18653/v1/N19-1423}{Association for Computational
  Linguistics}, \href{https://doi.org/10.18653/v1/N19-1423}{2019}.
  \href{https://doi.org/10.18653/v1/N19-1423}
{doi: {{%
10\hspace{.1pt}\discretionary{.}{%
}{.}\hspace{.4pt}18653\discretionary{/}{%
}{/}v1\discretionary{/}{%
}{/}N19\discretionary{%
}{-}{-}1423}}}


\bibitem{goffin2020interaction}
\href{https://doi.org/10.1145/3313831.3376842}{P.~Goffin, P.~Isenberg,
  T.~Blascheck, and W.~Willett}.
\newblock \href{https://doi.org/10.1145/3313831.3376842}{Interaction techniques
  for visual exploration using embedded word-scale visualizations}.
\newblock \href{https://doi.org/10.1145/3313831.3376842}{In {\em CHI 2020 -
  Conference on Human Factors in Computing Systems}}.
  \href{https://doi.org/10.1145/3313831.3376842}{ACM},
  \href{https://doi.org/10.1145/3313831.3376842}{2020}.
  \href{https://doi.org/10.1145/3313831.3376842}
{doi: {{%
10\hspace{.1pt}\discretionary{.}{%
}{.}\hspace{.4pt}1145\discretionary{/}{%
}{/}3313831\hspace{.1pt}\discretionary{.}{%
}{.}\hspace{.4pt}3376842}}}


\bibitem{gratzl2016visual}
\href{https://doi.org/10.1111/cgf.12925}{S.~Gratzl, A.~Lex, N.~Gehlenborg,
  N.~Cosgrove, and M.~Streit}.
\newblock \href{https://doi.org/10.1111/cgf.12925}{From visual exploration to
  storytelling and back again}.
\newblock \href{https://doi.org/10.1111/cgf.12925}{In {\em Computer Graphics
  Forum}}, \href{https://doi.org/10.1111/cgf.12925}{vol.~35},
  \href{https://doi.org/10.1111/cgf.12925}{pp. 491--500}.
  \href{https://doi.org/10.1111/cgf.12925}{Wiley Online Library},
  \href{https://doi.org/10.1111/cgf.12925}{2016}.
  \href{https://doi.org/10.1111/cgf.12925}
{doi: {{%
10\hspace{.1pt}\discretionary{.}{%
}{.}\hspace{.4pt}1111\discretionary{/}{%
}{/}cgf\hspace{.1pt}\discretionary{.}{%
}{.}\hspace{.4pt}12925}}}


\bibitem{heer2008generalized}
J.~Heer, M.~Agrawala, and W.~Willett.
\newblock Generalized selection via interactive query relaxation.
\newblock In {\em Proceedings of the SIGCHI Conference on Human Factors in
  Computing Systems}, pp. 959--968, 2008.

\bibitem{riche2018data}
N.~Henry~Riche, C.~Hurter, N.~Diakopoulos, and S.~Carpendale.
\newblock {\em Data-driven storytelling}.
\newblock CRC Press, 2018.

\bibitem{kim2020answering}
\href{https://doi.org/10.1145/3313831.3376467}{D.~H. Kim, E.~Hoque, and
  M.~Agrawala}.
\newblock \href{https://doi.org/10.1145/3313831.3376467}{Answering questions
  about charts and generating visual explanations}.
\newblock \href{https://doi.org/10.1145/3313831.3376467}{In {\em Proceedings of
  the 2020 CHI Conference on Human Factors in Computing Systems}},
  \href{https://doi.org/10.1145/3313831.3376467}{p. 1–13},
  \href{https://doi.org/10.1145/3313831.3376467}{2020}.
  \href{https://doi.org/10.1145/3313831.3376467}
{doi: {{%
10\hspace{.1pt}\discretionary{.}{%
}{.}\hspace{.4pt}1145\discretionary{/}{%
}{/}3313831\hspace{.1pt}\discretionary{.}{%
}{.}\hspace{.4pt}3376467}}}


\bibitem{kim2018facilitating}
D.~H. Kim, E.~Hoque, J.~Kim, and M.~Agrawala.
\newblock Facilitating document reading by linking text and tables.
\newblock In {\em Proceedings of the 31st Annual ACM Symposium on User
  Interface Software and Technology}, pp. 423--434, 2018.

\bibitem{kim2019datatoon}
\href{https://doi.org/10.1145/3290605.3300335}{N.~W. Kim, N.~Henry~Riche,
  B.~Bach, G.~Xu, M.~Brehmer, K.~Hinckley, M.~Pahud, H.~Xia, M.~J. McGuffin,
  and H.~Pfister}.
\newblock \href{https://doi.org/10.1145/3290605.3300335}{Datatoon: Drawing
  dynamic network comics with pen+ touch interaction}.
\newblock \href{https://doi.org/10.1145/3290605.3300335}{In {\em Proceedings of
  the 2019 CHI Conference on Human Factors in Computing Systems}},
  \href{https://doi.org/10.1145/3290605.3300335}{p. 105}.
  \href{https://doi.org/10.1145/3290605.3300335}{ACM},
  \href{https://doi.org/10.1145/3290605.3300335}{2019}.
  \href{https://doi.org/10.1145/3290605.3300335}
{doi: {{%
10\hspace{.1pt}\discretionary{.}{%
}{.}\hspace{.4pt}1145\discretionary{/}{%
}{/}3290605\hspace{.1pt}\discretionary{.}{%
}{.}\hspace{.4pt}3300335}}}


\bibitem{kim2019dataselfie}
\href{https://doi.org/10.1145/3290605.3300309}{N.~W. Kim, H.~Im,
  N.~Henry~Riche, A.~Wang, K.~Gajos, and H.~Pfister}.
\newblock \href{https://doi.org/10.1145/3290605.3300309}{{DataSelfie}:
  Empowering people to design personalized visuals to represent their data}.
\newblock \href{https://doi.org/10.1145/3290605.3300309}{In {\em Proceedings of
  the 2019 CHI Conference on Human Factors in Computing Systems}},
  \href{https://doi.org/10.1145/3290605.3300309}{p.~79},
  \href{https://doi.org/10.1145/3290605.3300309}{2019}.
  \href{https://doi.org/10.1145/3290605.3300309}
{doi: {{%
10\hspace{.1pt}\discretionary{.}{%
}{.}\hspace{.4pt}1145\discretionary{/}{%
}{/}3290605\hspace{.1pt}\discretionary{.}{%
}{.}\hspace{.4pt}3300309}}}


\bibitem{kim2016data}
\href{https://doi.org/10.1109/TVCG.2016.2598620}{N.~W. Kim, E.~Schweickart,
  Z.~Liu, M.~Dontcheva, W.~Li, J.~Popovic, and H.~Pfister}.
\newblock \href{https://doi.org/10.1109/TVCG.2016.2598620}{Data-driven guides:
  Supporting expressive design for information graphics}.
\newblock \href{https://doi.org/10.1109/TVCG.2016.2598620}{{\em IEEE
  Transactions on Visualization and Computer Graphics}},
  \href{https://doi.org/10.1109/TVCG.2016.2598620}{23(1):491--500},
  \href{https://doi.org/10.1109/TVCG.2016.2598620}{2016}.
  \href{https://doi.org/10.1109/TVCG.2016.2598620}
{doi: {{%
10\hspace{.1pt}\discretionary{.}{%
}{.}\hspace{.4pt}1109\discretionary{/}{%
}{/}TVCG\hspace{.1pt}\discretionary{.}{%
}{.}\hspace{.4pt}2016\hspace{.1pt}\discretionary{.}{%
}{.}\hspace{.4pt}2598620}}}


\bibitem{examplearticle}
R.~Kochhar.
\newblock Middle class fortunes in western europe.
\newblock
  \url{https://www.pewresearch.org/global/2017/04/24/middle-class-fortunes-in-western-europe/}.
\newblock Accessed: 2020-08-10.

\bibitem{kong2014extracting}
\href{https://doi.org/10.1145/2556288.2557241}{N.~Kong, M.~A. Hearst, and
  M.~Agrawala}.
\newblock \href{https://doi.org/10.1145/2556288.2557241}{Extracting references
  between text and charts via crowdsourcing}.
\newblock \href{https://doi.org/10.1145/2556288.2557241}{In {\em Proceedings of
  the SIGCHI conference on Human Factors in Computing Systems}},
  \href{https://doi.org/10.1145/2556288.2557241}{pp. 31--40}.
  \href{https://doi.org/10.1145/2556288.2557241}{ACM},
  \href{https://doi.org/10.1145/2556288.2557241}{2014}.
  \href{https://doi.org/10.1145/2556288.2557241}
{doi: {{%
10\hspace{.1pt}\discretionary{.}{%
}{.}\hspace{.4pt}1145\discretionary{/}{%
}{/}2556288\hspace{.1pt}\discretionary{.}{%
}{.}\hspace{.4pt}2557241}}}


\bibitem{kosara2013storytelling}
\href{https://doi.org/10.1109/MC.2013.36}{R.~Kosara and J.~Mackinlay}.
\newblock \href{https://doi.org/10.1109/MC.2013.36}{Storytelling: The next step
  for visualization}.
\newblock \href{https://doi.org/10.1109/MC.2013.36}{{\em Computer}},
  \href{https://doi.org/10.1109/MC.2013.36}{46(5):44--50},
  \href{https://doi.org/10.1109/MC.2013.36}{2013}.
  \href{https://doi.org/10.1109/MC.2013.36}
{doi: {{%
10\hspace{.1pt}\discretionary{.}{%
}{.}\hspace{.4pt}1109\discretionary{/}{%
}{/}MC\hspace{.1pt}\discretionary{.}{%
}{.}\hspace{.4pt}2013\hspace{.1pt}\discretionary{.}{%
}{.}\hspace{.4pt}36}}}


\bibitem{kwon2014visjockey}
B.~C. Kwon, F.~Stoffel, D.~J{\"a}ckle, B.~Lee, and D.~Keim.
\newblock {VisJockey}: Enriching data stories through orchestrated interactive
  visualization.
\newblock In {\em Poster Compendium of the Computation+ Journalism Symposium},
  vol.~3, p.~3, 2014.

\bibitem{lai2020automatic}
\href{https://doi.org/10.1145/3313831.3376443}{C.~Lai, Z.~Lin, R.~Jiang,
  Y.~Han, C.~Liu, and X.~Yuan}.
\newblock \href{https://doi.org/10.1145/3313831.3376443}{Automatic annotation
  synchronizing with textual description for visualization}.
\newblock \href{https://doi.org/10.1145/3313831.3376443}{In {\em Proceedings of
  the 2020 CHI Conference on Human Factors in Computing Systems}},
  \href{https://doi.org/10.1145/3313831.3376443}{pp. 1--13}.
  \href{https://doi.org/10.1145/3313831.3376443}{ACM},
  \href{https://doi.org/10.1145/3313831.3376443}{2020}.
  \href{https://doi.org/10.1145/3313831.3376443}
{doi: {{%
10\hspace{.1pt}\discretionary{.}{%
}{.}\hspace{.4pt}1145\discretionary{/}{%
}{/}3313831\hspace{.1pt}\discretionary{.}{%
}{.}\hspace{.4pt}3376443}}}


\bibitem{lalle2019gaze}
\href{https://doi.org/10.1109/TVCG.2019.2958540}{S.~Lall{\'e}, D.~Toker, and
  C.~Conati}.
\newblock \href{https://doi.org/10.1109/TVCG.2019.2958540}{Gaze-driven adaptive
  interventions for magazine-style narrative visualizations}.
\newblock \href{https://doi.org/10.1109/TVCG.2019.2958540}{{\em IEEE
  Transactions on Visualization and Computer Graphics}},
  \href{https://doi.org/10.1109/TVCG.2019.2958540}{2019}.
  \href{https://doi.org/10.1109/TVCG.2019.2958540}
{doi: {{%
10\hspace{.1pt}\discretionary{.}{%
}{.}\hspace{.4pt}1109\discretionary{/}{%
}{/}TVCG\hspace{.1pt}\discretionary{.}{%
}{.}\hspace{.4pt}2019\hspace{.1pt}\discretionary{.}{%
}{.}\hspace{.4pt}2958540}}}


\bibitem{latif2019vis}
\href{https://doi.org/10.1109/TVCG.2018.2865022}{S.~{Latif} and F.~{Beck}}.
\newblock \href{https://doi.org/10.1109/TVCG.2018.2865022}{{VIS} {A}uthor
  {P}rofiles: Interactive descriptions of publication records combining text
  and visualization}.
\newblock \href{https://doi.org/10.1109/TVCG.2018.2865022}{{\em IEEE
  Transactions on Visualization and Computer Graphics}},
  \href{https://doi.org/10.1109/TVCG.2018.2865022}{25(1):152--161},
  \href{https://doi.org/10.1109/TVCG.2018.2865022}{2019}.
  \href{https://doi.org/10.1109/TVCG.2018.2865022}
{doi: {{%
10\hspace{.1pt}\discretionary{.}{%
}{.}\hspace{.4pt}1109\discretionary{/}{%
}{/}TVCG\hspace{.1pt}\discretionary{.}{%
}{.}\hspace{.4pt}2018\hspace{.1pt}\discretionary{.}{%
}{.}\hspace{.4pt}2865022}}}


\bibitem{latif2021deeper}
\href{https://doi.org/10.1111/cgf.14309}{S.~Latif, S.~Chen, and F.~Beck}.
\newblock \href{https://doi.org/10.1111/cgf.14309}{A deeper understanding of
  visualization-text interplay in geographic data-driven stories}.
\newblock \href{https://doi.org/10.1111/cgf.14309}{{\em Computer Graphics
  Forum}}, \href{https://doi.org/10.1111/cgf.14309}{40(3):311--322},
  \href{https://doi.org/10.1111/cgf.14309}{2021}.
  \href{https://doi.org/10.1111/cgf.14309}
{doi: {{%
10\hspace{.1pt}\discretionary{.}{%
}{.}\hspace{.4pt}1111\discretionary{/}{%
}{/}cgf\hspace{.1pt}\discretionary{.}{%
}{.}\hspace{.4pt}14309}}}


\bibitem{latif2018exploring}
\href{https://doi.org/10.2312/eurovisshort.20181084}{S.~Latif, D.~Liu, and
  F.~Beck}.
\newblock \href{https://doi.org/10.2312/eurovisshort.20181084}{{Exploring
  interactive linking between text and visualization}}.
\newblock \href{https://doi.org/10.2312/eurovisshort.20181084}{In {\em EuroVis
  2018 - Short Papers}}.
  \href{https://doi.org/10.2312/eurovisshort.20181084}{The Eurographics
  Association}, \href{https://doi.org/10.2312/eurovisshort.20181084}{2018}.
  \href{https://doi.org/10.2312/eurovisshort.20181084}
{doi: {{%
10\hspace{.1pt}\discretionary{.}{%
}{.}\hspace{.4pt}2312\discretionary{/}{%
}{/}eurovisshort\hspace{.1pt}\discretionary{.}{%
}{.}\hspace{.4pt}20181084}}}


\bibitem{latif2019authoring}
\href{https://doi.org/doi:10.2312/evs.20191180}{S.~Latif, K.~Su, and F.~Beck}.
\newblock \href{https://doi.org/doi:10.2312/evs.20191180}{Authoring combined
  textual and visual descriptions of graph data}.
\newblock \href{https://doi.org/doi:10.2312/evs.20191180}{In {\em EuroVis 2019
  - Short Papers}}. \href{https://doi.org/doi:10.2312/evs.20191180}{The
  Eurographics Association},
  \href{https://doi.org/doi:10.2312/evs.20191180}{may 2019}.
  \href{https://doi.org/doi:10.2312/evs.20191180}
{doi: {{%
doi\discretionary{:}{%
}{:}10\hspace{.1pt}\discretionary{.}{%
}{.}\hspace{.4pt}2312\discretionary{/}{%
}{/}evs\hspace{.1pt}\discretionary{.}{%
}{.}\hspace{.4pt}20191180}}}


\bibitem{Latif2021talking}
\href{https://doi.org/10.1109/MCG.2021.3058129}{S.~{Latif}, H.~{Tarner}, and
  F.~{Beck}}.
\newblock \href{https://doi.org/10.1109/MCG.2021.3058129}{Talking realities:
  Audio guides in virtual reality visualizations}.
\newblock \href{https://doi.org/10.1109/MCG.2021.3058129}{{\em IEEE Computer
  Graphics and Applications}},
  \href{https://doi.org/10.1109/MCG.2021.3058129}{pp. 1--1},
  \href{https://doi.org/10.1109/MCG.2021.3058129}{2021}.
  \href{https://doi.org/10.1109/MCG.2021.3058129}
{doi: {{%
10\hspace{.1pt}\discretionary{.}{%
}{.}\hspace{.4pt}1109\discretionary{/}{%
}{/}MCG\hspace{.1pt}\discretionary{.}{%
}{.}\hspace{.4pt}2021\hspace{.1pt}\discretionary{.}{%
}{.}\hspace{.4pt}3058129}}}


\bibitem{lee2015more}
\href{https://doi.org/10.1109/MCG.2015.99}{B.~Lee, N.~Henry~Riche, P.~Isenberg,
  and S.~Carpendale}.
\newblock \href{https://doi.org/10.1109/MCG.2015.99}{More than telling a story:
  Transforming data into visually shared stories}.
\newblock \href{https://doi.org/10.1109/MCG.2015.99}{{\em IEEE Computer
  Graphics and Applications}},
  \href{https://doi.org/10.1109/MCG.2015.99}{35(5):84--90},
  \href{https://doi.org/10.1109/MCG.2015.99}{2015}.
  \href{https://doi.org/10.1109/MCG.2015.99}
{doi: {{%
10\hspace{.1pt}\discretionary{.}{%
}{.}\hspace{.4pt}1109\discretionary{/}{%
}{/}MCG\hspace{.1pt}\discretionary{.}{%
}{.}\hspace{.4pt}2015\hspace{.1pt}\discretionary{.}{%
}{.}\hspace{.4pt}99}}}


\bibitem{liu2018data}
\href{https://doi.org/10.1145/3173574.3173697}{Z.~Liu, J.~Thompson, A.~Wilson,
  M.~Dontcheva, J.~Delorey, S.~Grigg, B.~Kerr, and J.~Stasko}.
\newblock \href{https://doi.org/10.1145/3173574.3173697}{Data illustrator:
  Augmenting vector design tools with lazy data binding for expressive
  visualization authoring}.
\newblock \href{https://doi.org/10.1145/3173574.3173697}{In {\em Proceedings of
  the 2018 CHI Conference on Human Factors in Computing Systems}},
  \href{https://doi.org/10.1145/3173574.3173697}{p. 123}.
  \href{https://doi.org/10.1145/3173574.3173697}{ACM},
  \href{https://doi.org/10.1145/3173574.3173697}{2018}.
  \href{https://doi.org/10.1145/3173574.3173697}
{doi: {{%
10\hspace{.1pt}\discretionary{.}{%
}{.}\hspace{.4pt}1145\discretionary{/}{%
}{/}3173574\hspace{.1pt}\discretionary{.}{%
}{.}\hspace{.4pt}3173697}}}


\bibitem{mackinlay1986automating}
\href{https://doi.org/10.1145/22949.22950}{J.~Mackinlay}.
\newblock \href{https://doi.org/10.1145/22949.22950}{Automating the design of
  graphical presentations of relational information}.
\newblock \href{https://doi.org/10.1145/22949.22950}{{\em {ACM} Transactions On
  Graphics}}, \href{https://doi.org/10.1145/22949.22950}{5(2):110--141},
  \href{https://doi.org/10.1145/22949.22950}{1986}.
  \href{https://doi.org/10.1145/22949.22950}
{doi: {{%
10\hspace{.1pt}\discretionary{.}{%
}{.}\hspace{.4pt}1145\discretionary{/}{%
}{/}22949\hspace{.1pt}\discretionary{.}{%
}{.}\hspace{.4pt}22950}}}


\bibitem{mayer201412}
R.~E. Mayer and L.~Fiorella.
\newblock 12 principles for reducing extraneous processing in multimedia
  learning: Coherence, signaling, redundancy, spatial contiguity, and temporal
  contiguity principles.
\newblock In {\em The Cambridge handbook of multimedia learning}, vol. 279.
  Cambridge University Press, 2014.

\bibitem{mcneill2019viz}
\href{https://doi.org/10.2312/evs.20191177}{G.~McNeill and S.~A. Hale}.
\newblock \href{https://doi.org/10.2312/evs.20191177}{{Viz-Blocks: Building
  Visualizations and Documents in the Browser}}.
\newblock \href{https://doi.org/10.2312/evs.20191177}{In J.~Johansson,
  F.~Sadlo, and G.~E. Marai, eds., {\em EuroVis 2019 - Short Papers}}.
  \href{https://doi.org/10.2312/evs.20191177}{The Eurographics Association},
  \href{https://doi.org/10.2312/evs.20191177}{2019}.
  \href{https://doi.org/10.2312/evs.20191177}
{doi: {{%
10\hspace{.1pt}\discretionary{.}{%
}{.}\hspace{.4pt}2312\discretionary{/}{%
}{/}evs\hspace{.1pt}\discretionary{.}{%
}{.}\hspace{.4pt}20191177}}}


\bibitem{mikolov2018advances}
T.~Mikolov, E.~Grave, P.~Bojanowski, C.~Puhrsch, and A.~Joulin.
\newblock Advances in pre-training distributed word representations.
\newblock In {\em Proceedings of the International Conference on Language
  Resources and Evaluation (LREC 2018)}, 2018.

\bibitem{NIPS2013_9aa42b31}
\href{https://proceedings.neurips.cc/paper/2013/file/9aa42b31882ec039965f3c4923ce901b-Paper.pdf}{T.~Mikolov,
  I.~Sutskever, K.~Chen, G.~S. Corrado, and J.~Dean}.
\newblock
  \href{https://proceedings.neurips.cc/paper/2013/file/9aa42b31882ec039965f3c4923ce901b-Paper.pdf}{Distributed
  representations of words and phrases and their compositionality}.
\newblock
  \href{https://proceedings.neurips.cc/paper/2013/file/9aa42b31882ec039965f3c4923ce901b-Paper.pdf}{In
  {\em Advances in Neural Information Processing Systems}},
  \href{https://proceedings.neurips.cc/paper/2013/file/9aa42b31882ec039965f3c4923ce901b-Paper.pdf}{vol.~26}.
  \href{https://proceedings.neurips.cc/paper/2013/file/9aa42b31882ec039965f3c4923ce901b-Paper.pdf}{Curran
  Associates, Inc.},
  \href{https://proceedings.neurips.cc/paper/2013/file/9aa42b31882ec039965f3c4923ce901b-Paper.pdf}{2013}.

\bibitem{elsevier}
P.~Nowakowski, E.~Ciepiela, D.~Har{\k{e}}{\.z}lak, J.~Kocot, M.~Kasztelnik,
  T.~Barty{\'n}ski, J.~Meizner, G.~Dyk, and M.~Malawski.
\newblock The collage authoring environment.
\newblock {\em Procedia Computer Science}, 4:608--617, 2011.

\bibitem{ottley2019curious}
\href{https://doi.org/10.2312/evs.20191181}{A.~Ottley, A.~Kaszowska, R.~J.
  Crouser, and E.~M. Peck}.
\newblock \href{https://doi.org/10.2312/evs.20191181}{{The Curious Case of
  Combining Text and Visualization}}.
\newblock \href{https://doi.org/10.2312/evs.20191181}{In J.~Johansson,
  F.~Sadlo, and G.~E. Marai, eds., {\em EuroVis 2019 - Short Papers}}.
  \href{https://doi.org/10.2312/evs.20191181}{The Eurographics Association},
  \href{https://doi.org/10.2312/evs.20191181}{2019}.
  \href{https://doi.org/10.2312/evs.20191181}
{doi: {{%
10\hspace{.1pt}\discretionary{.}{%
}{.}\hspace{.4pt}2312\discretionary{/}{%
}{/}evs\hspace{.1pt}\discretionary{.}{%
}{.}\hspace{.4pt}20191181}}}


\bibitem{sweller2005implications}
F.~Paas and J.~Sweller.
\newblock Implications of cognitive load theory for multimedia learning.
\newblock {\em The Cambridge handbook of multimedia learning}, 27:27--42, 2014.

\bibitem{ren2017chartaccent}
\href{https://doi.org/10.1109/pacificvis.2017.8031599}{D.~Ren, M.~Brehmer,
  B.~Lee, T.~H{\"o}llerer, and E.~K. Choe}.
\newblock \href{https://doi.org/10.1109/pacificvis.2017.8031599}{{ChartAccent}:
  Annotation for data-driven storytelling}.
\newblock \href{https://doi.org/10.1109/pacificvis.2017.8031599}{In {\em 2017
  IEEE Pacific Visualization Symposium (PacificVis)}},
  \href{https://doi.org/10.1109/pacificvis.2017.8031599}{pp. 230--239}.
  \href{https://doi.org/10.1109/pacificvis.2017.8031599}{IEEE},
  \href{https://doi.org/10.1109/pacificvis.2017.8031599}{2017}.
  \href{https://doi.org/10.1109/pacificvis.2017.8031599}
{doi: {{%
10\hspace{.1pt}\discretionary{.}{%
}{.}\hspace{.4pt}1109\discretionary{/}{%
}{/}pacificvis\hspace{.1pt}\discretionary{.}{%
}{.}\hspace{.4pt}2017\hspace{.1pt}\discretionary{.}{%
}{.}\hspace{.4pt}8031599}}}


\bibitem{ren2014ivisdesigner}
\href{https://doi.org/10.1109/TVCG.2014.2346291}{D.~Ren, T.~H{\"o}llerer, and
  X.~Yuan}.
\newblock \href{https://doi.org/10.1109/TVCG.2014.2346291}{{iVisDesigner}:
  Expressive interactive design of information visualizations}.
\newblock \href{https://doi.org/10.1109/TVCG.2014.2346291}{{\em IEEE
  Transactions on Visualization and Computer Graphics}},
  \href{https://doi.org/10.1109/TVCG.2014.2346291}{20(12):2092--2101},
  \href{https://doi.org/10.1109/TVCG.2014.2346291}{2014}.
  \href{https://doi.org/10.1109/TVCG.2014.2346291}
{doi: {{%
10\hspace{.1pt}\discretionary{.}{%
}{.}\hspace{.4pt}1109\discretionary{/}{%
}{/}TVCG\hspace{.1pt}\discretionary{.}{%
}{.}\hspace{.4pt}2014\hspace{.1pt}\discretionary{.}{%
}{.}\hspace{.4pt}2346291}}}


\bibitem{ren2018charticulator}
\href{https://doi.org/10.1109/TVCG.2018.2865158}{D.~Ren, B.~Lee, and
  M.~Brehmer}.
\newblock \href{https://doi.org/10.1109/TVCG.2018.2865158}{Charticulator:
  Interactive construction of bespoke chart layouts}.
\newblock \href{https://doi.org/10.1109/TVCG.2018.2865158}{{\em IEEE
  Transactions on Visualization and Computer Graphics}},
  \href{https://doi.org/10.1109/TVCG.2018.2865158}{25(1):789--799},
  \href{https://doi.org/10.1109/TVCG.2018.2865158}{2018}.
  \href{https://doi.org/10.1109/TVCG.2018.2865158}
{doi: {{%
10\hspace{.1pt}\discretionary{.}{%
}{.}\hspace{.4pt}1109\discretionary{/}{%
}{/}TVCG\hspace{.1pt}\discretionary{.}{%
}{.}\hspace{.4pt}2018\hspace{.1pt}\discretionary{.}{%
}{.}\hspace{.4pt}2865158}}}


\bibitem{satyanarayan2014authoring}
\href{https://doi.org/10.1111/cgf.12392}{A.~Satyanarayan and J.~Heer}.
\newblock \href{https://doi.org/10.1111/cgf.12392}{Authoring narrative
  visualizations with {Ellipsis}}.
\newblock \href{https://doi.org/10.1111/cgf.12392}{In {\em Computer Graphics
  Forum}}, \href{https://doi.org/10.1111/cgf.12392}{vol.~33},
  \href{https://doi.org/10.1111/cgf.12392}{pp. 361--370}.
  \href{https://doi.org/10.1111/cgf.12392}{Wiley Online Library},
  \href{https://doi.org/10.1111/cgf.12392}{2014}.
  \href{https://doi.org/10.1111/cgf.12392}
{doi: {{%
10\hspace{.1pt}\discretionary{.}{%
}{.}\hspace{.4pt}1111\discretionary{/}{%
}{/}cgf\hspace{.1pt}\discretionary{.}{%
}{.}\hspace{.4pt}12392}}}


\bibitem{satyanarayan2014lyra}
\href{https://doi.org/10.1111/cgf.12391}{A.~Satyanarayan and J.~Heer}.
\newblock \href{https://doi.org/10.1111/cgf.12391}{Lyra: An interactive
  visualization design environment}.
\newblock \href{https://doi.org/10.1111/cgf.12391}{vol.~33},
  \href{https://doi.org/10.1111/cgf.12391}{pp. 351--360},
  \href{https://doi.org/10.1111/cgf.12391}{2014}.
  \href{https://doi.org/10.1111/cgf.12391}
{doi: {{%
10\hspace{.1pt}\discretionary{.}{%
}{.}\hspace{.4pt}1111\discretionary{/}{%
}{/}cgf\hspace{.1pt}\discretionary{.}{%
}{.}\hspace{.4pt}12391}}}


\bibitem{satyanarayan2017vega-lite}
\href{https://doi.org/10.1109/TVCG.2016.2599030}{A.~Satyanarayan, D.~Moritz,
  K.~Wongsuphasawat, and J.~Heer}.
\newblock \href{https://doi.org/10.1109/TVCG.2016.2599030}{Vega-lite: A grammar
  of interactive graphics}.
\newblock \href{https://doi.org/10.1109/TVCG.2016.2599030}{{\em IEEE
  Transactions on Visualization and Computer Graphics (Proc. InfoVis)}},
  \href{https://doi.org/10.1109/TVCG.2016.2599030}{2017}.
  \href{https://doi.org/10.1109/TVCG.2016.2599030}
{doi: {{%
10\hspace{.1pt}\discretionary{.}{%
}{.}\hspace{.4pt}1109\discretionary{/}{%
}{/}TVCG\hspace{.1pt}\discretionary{.}{%
}{.}\hspace{.4pt}2016\hspace{.1pt}\discretionary{.}{%
}{.}\hspace{.4pt}2599030}}}


\bibitem{satyanarayan2016vega}
\href{https://doi.org/10.1109/TVCG.2015.2467091}{A.~{Satyanarayan},
  R.~{Russell}, J.~{Hoffswell}, and J.~{Heer}}.
\newblock \href{https://doi.org/10.1109/TVCG.2015.2467091}{Reactive vega: A
  streaming dataflow architecture for declarative interactive visualization}.
\newblock \href{https://doi.org/10.1109/TVCG.2015.2467091}{{\em IEEE
  Transactions on Visualization and Computer Graphics}},
  \href{https://doi.org/10.1109/TVCG.2015.2467091}{22(1):659--668},
  \href{https://doi.org/10.1109/TVCG.2015.2467091}{2016}.
  \href{https://doi.org/10.1109/TVCG.2015.2467091}
{doi: {{%
10\hspace{.1pt}\discretionary{.}{%
}{.}\hspace{.4pt}1109\discretionary{/}{%
}{/}TVCG\hspace{.1pt}\discretionary{.}{%
}{.}\hspace{.4pt}2015\hspace{.1pt}\discretionary{.}{%
}{.}\hspace{.4pt}2467091}}}


\bibitem{segel2010narrative}
\href{https://doi.org/10.1109/tvcg.2010.179}{E.~Segel and J.~Heer}.
\newblock \href{https://doi.org/10.1109/tvcg.2010.179}{Narrative visualization:
  Telling stories with data}.
\newblock \href{https://doi.org/10.1109/tvcg.2010.179}{{\em IEEE Transactions
  on Visualization and Computer Graphics}},
  \href{https://doi.org/10.1109/tvcg.2010.179}{16(6):1139--1148},
  \href{https://doi.org/10.1109/tvcg.2010.179}{2010}.
  \href{https://doi.org/10.1109/tvcg.2010.179}
{doi: {{%
10\hspace{.1pt}\discretionary{.}{%
}{.}\hspace{.4pt}1109\discretionary{/}{%
}{/}tvcg\hspace{.1pt}\discretionary{.}{%
}{.}\hspace{.4pt}2010\hspace{.1pt}\discretionary{.}{%
}{.}\hspace{.4pt}179}}}


\bibitem{srinivasan2018augmenting}
\href{https://doi.org/10.1109/TVCG.2018.2865145}{A.~Srinivasan, S.~M. Drucker,
  A.~Endert, and J.~Stasko}.
\newblock \href{https://doi.org/10.1109/TVCG.2018.2865145}{Augmenting
  visualizations with interactive data facts to facilitate interpretation and
  communication}.
\newblock \href{https://doi.org/10.1109/TVCG.2018.2865145}{{\em IEEE
  Transactions on Visualization and Computer Graphics}},
  \href{https://doi.org/10.1109/TVCG.2018.2865145}{25(1):672--681},
  \href{https://doi.org/10.1109/TVCG.2018.2865145}{2018}.
  \href{https://doi.org/10.1109/TVCG.2018.2865145}
{doi: {{%
10\hspace{.1pt}\discretionary{.}{%
}{.}\hspace{.4pt}1109\discretionary{/}{%
}{/}TVCG\hspace{.1pt}\discretionary{.}{%
}{.}\hspace{.4pt}2018\hspace{.1pt}\discretionary{.}{%
}{.}\hspace{.4pt}2865145}}}


\bibitem{srinivasan2017natural}
\href{https://doi.org/10.2312/eurovisshort.20171133}{A.~Srinivasan and
  J.~Stasko}.
\newblock \href{https://doi.org/10.2312/eurovisshort.20171133}{Natural language
  interfaces for data analysis with visualization: Considering what has and
  could be asked}.
\newblock \href{https://doi.org/10.2312/eurovisshort.20171133}{In {\em
  Proceedings of the Eurographics/IEEE VGTC Conference on Visualization: Short
  Papers}}, \href{https://doi.org/10.2312/eurovisshort.20171133}{pp. 55--59},
  \href{https://doi.org/10.2312/eurovisshort.20171133}{2017}.
  \href{https://doi.org/10.2312/eurovisshort.20171133}
{doi: {{%
10\hspace{.1pt}\discretionary{.}{%
}{.}\hspace{.4pt}2312\discretionary{/}{%
}{/}eurovisshort\hspace{.1pt}\discretionary{.}{%
}{.}\hspace{.4pt}20171133}}}


\bibitem{steinberger2011context}
\href{https://doi.org/10.1109/TVCG.2011.183}{M.~Steinberger, M.~Waldner,
  M.~Streit, A.~Lex, and D.~Schmalstieg}.
\newblock \href{https://doi.org/10.1109/TVCG.2011.183}{Context-preserving
  visual links}.
\newblock \href{https://doi.org/10.1109/TVCG.2011.183}{{\em IEEE Transactions
  on Visualization and Computer Graphics}},
  \href{https://doi.org/10.1109/TVCG.2011.183}{17(12):2249--2258},
  \href{https://doi.org/10.1109/TVCG.2011.183}{2011}.
  \href{https://doi.org/10.1109/TVCG.2011.183}
{doi: {{%
10\hspace{.1pt}\discretionary{.}{%
}{.}\hspace{.4pt}1109\discretionary{/}{%
}{/}TVCG\hspace{.1pt}\discretionary{.}{%
}{.}\hspace{.4pt}2011\hspace{.1pt}\discretionary{.}{%
}{.}\hspace{.4pt}183}}}


\bibitem{stolte2002polaris}
\href{https://doi.org/10.1109/2945.981851}{C.~Stolte, D.~Tang, and
  P.~Hanrahan}.
\newblock \href{https://doi.org/10.1109/2945.981851}{Polaris: A system for
  query, analysis, and visualization of multidimensional relational databases}.
\newblock \href{https://doi.org/10.1109/2945.981851}{{\em IEEE Transactions on
  Visualization and Computer Graphics}},
  \href{https://doi.org/10.1109/2945.981851}{8(1):52--65},
  \href{https://doi.org/10.1109/2945.981851}{2002}.
  \href{https://doi.org/10.1109/2945.981851}
{doi: {{%
10\hspace{.1pt}\discretionary{.}{%
}{.}\hspace{.4pt}1109\discretionary{/}{%
}{/}2945\hspace{.1pt}\discretionary{.}{%
}{.}\hspace{.4pt}981851}}}


\bibitem{sultanum2021leveraging}
\href{https://doi.org/10.1145/3411764.3445354}{N.~Sultanum, F.~Chevalier,
  Z.~Bylinskii, and Z.~Liu}.
\newblock \href{https://doi.org/10.1145/3411764.3445354}{Leveraging text-chart
  links to support authoring of data-driven articles with vizflow}.
\newblock \href{https://doi.org/10.1145/3411764.3445354}{In {\em Proceedings of
  the 2021 CHI Conference on Human Factors in Computing Systems}},
  \href{https://doi.org/10.1145/3411764.3445354}{pp. 1--17},
  \href{https://doi.org/10.1145/3411764.3445354}{2021}.
  \href{https://doi.org/10.1145/3411764.3445354}
{doi: {{%
10\hspace{.1pt}\discretionary{.}{%
}{.}\hspace{.4pt}1145\discretionary{/}{%
}{/}3411764\hspace{.1pt}\discretionary{.}{%
}{.}\hspace{.4pt}3445354}}}


\bibitem{economist2020where}
{The Economist}.
\newblock Where are the world’s best english-speakers?
\newblock
  \url{https://www.economist.com/graphic-detail/2019/12/04/where-are-the-worlds-best-english-speakers},
  2019.
\newblock Accessed: 2020-09-17.

\bibitem{economist2020higher}
{The Economist}.
\newblock Higher minimum wages are linked to lower suicide rate.
\newblock
  \url{https://www.economist.com/graphic-detail/2020/01/20/higher-minimum-wages-are-linked-to-lower-suicide-rates},
  2020.
\newblock Accessed: 2020-09-17.

\bibitem{economist2020richest}
{The Economist}.
\newblock The richest countries have the fewest road deaths.
\newblock
  \url{https://www.economist.com/graphic-detail/2020/01/27/the-richest-countries-have-the-fewest-road-deaths},
  2020.
\newblock Accessed: 2020-09-17.

\bibitem{wiseman-etal-2017-challenges}
\href{https://doi.org/10.18653/v1/D17-1239}{S.~Wiseman, S.~Shieber, and
  A.~Rush}.
\newblock \href{https://doi.org/10.18653/v1/D17-1239}{Challenges in
  data-to-document generation}.
\newblock \href{https://doi.org/10.18653/v1/D17-1239}{In {\em Proceedings of
  the 2017 Conference on Empirical Methods in Natural Language Processing}},
  \href{https://doi.org/10.18653/v1/D17-1239}{pp. 2253--2263},
  \href{https://doi.org/10.18653/v1/D17-1239}{2017}.
  \href{https://doi.org/10.18653/v1/D17-1239}
{doi: {{%
10\hspace{.1pt}\discretionary{.}{%
}{.}\hspace{.4pt}18653\discretionary{/}{%
}{/}v1\discretionary{/}{%
}{/}D17\discretionary{%
}{-}{-}1239}}}


\bibitem{wongsuphasawat2015voyager}
\href{https://doi.org/10.1109/TVCG.2015.2467191}{K.~Wongsuphasawat, D.~Moritz,
  A.~Anand, J.~Mackinlay, B.~Howe, and J.~Heer}.
\newblock \href{https://doi.org/10.1109/TVCG.2015.2467191}{Voyager: Exploratory
  analysis via faceted browsing of visualization recommendations}.
\newblock \href{https://doi.org/10.1109/TVCG.2015.2467191}{{\em IEEE
  Transactions on Visualization and Computer Graphics}},
  \href{https://doi.org/10.1109/TVCG.2015.2467191}{22(1):649--658},
  \href{https://doi.org/10.1109/TVCG.2015.2467191}{2015}.
  \href{https://doi.org/10.1109/TVCG.2015.2467191}
{doi: {{%
10\hspace{.1pt}\discretionary{.}{%
}{.}\hspace{.4pt}1109\discretionary{/}{%
}{/}TVCG\hspace{.1pt}\discretionary{.}{%
}{.}\hspace{.4pt}2015\hspace{.1pt}\discretionary{.}{%
}{.}\hspace{.4pt}2467191}}}


\bibitem{xia2018dataink}
\href{https://doi.org/10.1145/3173574.3173797}{H.~Xia, N.~Henry~Riche,
  F.~Chevalier, B.~De~Araujo, and D.~Wigdor}.
\newblock \href{https://doi.org/10.1145/3173574.3173797}{{DataInk}: Direct and
  creative data-oriented drawing}.
\newblock \href{https://doi.org/10.1145/3173574.3173797}{In {\em Proceedings of
  the 2018 CHI Conference on Human Factors in Computing Systems}},
  \href{https://doi.org/10.1145/3173574.3173797}{p. 223}.
  \href{https://doi.org/10.1145/3173574.3173797}{ACM},
  \href{https://doi.org/10.1145/3173574.3173797}{2018}.
  \href{https://doi.org/10.1145/3173574.3173797}
{doi: {{%
10\hspace{.1pt}\discretionary{.}{%
}{.}\hspace{.4pt}1145\discretionary{/}{%
}{/}3173574\hspace{.1pt}\discretionary{.}{%
}{.}\hspace{.4pt}3173797}}}


\bibitem{zhi2019linking}
\href{https://doi.org/10.1111/cgf.13719}{Q.~Zhi, A.~Ottley, and R.~Metoyer}.
\newblock \href{https://doi.org/10.1111/cgf.13719}{Linking and layout:
  Exploring the integration of text and visualization in storytelling}.
\newblock \href{https://doi.org/10.1111/cgf.13719}{In {\em Computer Graphics
  Forum}}, \href{https://doi.org/10.1111/cgf.13719}{vol.~38},
  \href{https://doi.org/10.1111/cgf.13719}{pp. 675--685}.
  \href{https://doi.org/10.1111/cgf.13719}{Wiley Online Library},
  \href{https://doi.org/10.1111/cgf.13719}{2019}.
  \href{https://doi.org/10.1111/cgf.13719}
{doi: {{%
10\hspace{.1pt}\discretionary{.}{%
}{.}\hspace{.4pt}1111\discretionary{/}{%
}{/}cgf\hspace{.1pt}\discretionary{.}{%
}{.}\hspace{.4pt}13719}}}


\end{thebibliography}
